\documentclass[prd,singlecolumn,superscriptaddress]{revtex4}
\usepackage[utf8]{inputenc}
\usepackage{empheq}
\usepackage{amsmath}
\usepackage{amsfonts}
\usepackage{amssymb,bm,txfonts}
\usepackage{mathtools}
\usepackage{stmaryrd}
\usepackage{dsfont}
\usepackage{graphicx}
\usepackage[bookmarksnumbered=true]{hyperref}
\usepackage{stackrel}
\usepackage{color}
\usepackage{comment}
\usepackage{xcolor}
\usepackage{enumitem}
 \usepackage{xr}

\usepackage{graphicx}
\usepackage{siunitx}
\usepackage{fancyhdr}

\def \ee{\end{equation}}
\def \be{\begin{equation}}
\def \eea{\end{eqnarray}}
\def \bea{\begin{eqnarray}}

\begin{document}
\title{
Quantum Uncertainties of Static Spherically Symmetric Spacetimes
%Uncertainty relations of static spherically symmetric spacetime
}
\author{Benjamin Koch}
\email{benjamin.koch@tuwien.ac.at}
%\affiliation{Institut f\"ur Theoretische Physik und Atominstitut,
% Technische Universit\"at Wien,
% Wiedner Hauptstrasse 8--10,
% A-1040 Vienna, Austria}
\affiliation{Institut f\"ur Theoretische Physik, Technische Universit\"at Wien, Wiedner Hauptstrasse 8--10, A-1040 Vienna, Austria}
\affiliation{Atominstitut, Technische Universit\"at Wien,  Stadionallee 2, A-1020 Vienna, Austria}
 \affiliation{Pontificia Universidad Cat\'olica de Chile \\ Instituto de F\'isica, Pontificia Universidad Cat\'olica de Chile, \\
Casilla 306, Santiago, Chile}
\author{Ali Riahinia}
\email{ali.riahinia@tuwien.ac.at}
%\affiliation{Institut f\"ur Theoretische Physik und Atominstitut,
% Technische Universit\"at Wien,
% Wiedner Hauptstrasse 8--10,
% A-1040 Vienna, Austria}
\affiliation{Institut f\"ur Theoretische Physik, Technische Universit\"at Wien, Wiedner Hauptstrasse 8--10, A-1040 Vienna, Austria}
\affiliation{Atominstitut, Technische Universit\"at Wien,  Stadionallee 2, A-1020 Vienna, Austria}
% \affiliation{Pontificia Universidad Cat\'olica de Chile \\ Instituto de F\'isica, Pontificia Universidad Cat\'olica de Chile, \\
%Casilla 306, Santiago, Chile}
%
\begin{abstract}
We present a canonical quantization framework for static spherically symmetric spacetimes described by the Einstein–Hilbert action with a cosmological constant. In addition to recovering the classical Schwarzschild–(Anti)-de Sitter solutions via the Ehrenfest theorem, we investigate the quantum uncertainty relations that arise among the geometric operators in this setup. Our analysis uncovers an intriguing relation to black hole thermodynamics and opens a new angle towards generalized uncertainty relations. 
We further obtain an upper 
and a lower limit of the mass that is allowed in our model, for a given value of the cosmological constant.
Both limits, when evaluated for the known value of the cosmological constant, have a stunning relation to observed bounds.
These findings open a promising avenue for deeper insights into how quantum effects manifest in spacetime geometry and gravitational systems.
\end{abstract}
\maketitle
\tableofcontents
%%%%%%%%%%%%%%%%%%%%%%%
\section{Introduction}

The need for a quantum theory of gravity emerges from the long-standing foundational tension
between general relativity and quantum field theory.
General relativity describes gravitational phenomena with remarkable accuracy on macroscopic scales,
yet its classical description may not be adequate at the tiniest distances where quantum effects dominate.
Conversely, standard quantum theories, in their present form, rely on a fixed background geometry and
cannot be directly applied in settings where the space-time curvature itself becomes extreme,
for instance near black-hole singularities or at the Big Bang.
Reconciling these two cornerstones of modern physics in a single, self-consistent framework is the primary goal of quantum-gravity research, and it underpins a wide spectrum of programmes ranging from canonical and covariant quantisation to path-integral and other non-perturbative approaches.

%%%%%%%%%%%%%%%%%%%%%%%
\subsection{Canonical Quantum Gravity}

The canonical approach to quantum gravity rests on the pillars of the ADM (Arnowitt--Deser--Misner)  approach \cite{Arnowitt:1959ah,Arnowitt:1962hi}. (For a pedagogical introduction see \cite{Banados:2016zim,Thiemann:2001gmi}).
It proceeds by decomposing a four-dimensional spacetime $\mathcal{M}$ into a foliation of three-dimensional spatial hypersurfaces $\Sigma_t$ labeled by a time coordinate $t$. Concretely, the spacetime metric $g_{\mu \nu}$ is re-expressed in terms of
\be
ds^2 = -N^2 dt^2 + q_{ij}(dx^i + N^i dt)(dx^j + N^j dt),
\ee
where
\begin{itemize}
  \item The three-metric $q_{ij}$ on each hypersurface $\Sigma_t$.
  \item The \emph{lapse} function $N$, which governs how these hypersurfaces are separated in the temporal direction.
  \item The \emph{shift} vector $N^i$, which describes the ``horizontal'' displacement of spatial coordinates from one time slice to the next.
\end{itemize}

In this ADM formalism, the canonical variables are typically chosen as the three-metric $q_{ij}$ and its conjugate momentum $\pi^{ij}$ (related to the extrinsic curvature of $\Sigma_t$). The full dynamics are then encoded in the Hamiltonian and momentum (diffeomorphism) constraints:
\begin{equation}
\mathcal{H}(q_{ij}, \pi^{ij}) \approx 0, \quad \mathcal{H}_i(q_{ij}, \pi^{ij}) \approx 0,
\end{equation}
which reflect the invariances of general relativity under time reparametrizations and spatial diffeomorphisms, respectively. The choice of a distinct time slicing, signaled by the \emph{lapse} function $N$ and \emph{shift} vector $N^i$, is crucial. It effectively picks out a preferred (albeit dynamically determined and gauge-dependent) temporal evolution of the spatial geometry.

From this classical Hamiltonian structure, different paths to canonical quantum gravity diverge. The traditional approach, often termed quantum geometrodynamics, aims to promote $q_{ij}$ and $\pi^{ij}$ directly to operators acting on wave functionals $\Psi[q_{ij}]$. Imposing the Hamiltonian constraint at the quantum level leads to the Wheeler-DeWitt (WDW) equation~\cite{DeWitt:1967uc,DeWitt:1967ub,DeWitt:1967yk}:
\begin{equation}
\hat{\mathcal{H}} \, \Psi[q_{ij}] = 0.
\end{equation}
Defining this equation rigorously is a profound challenge, largely due to the singular nature of the operators involved and the difficulty in constructing a well-defined inner product on the space of three-geometries (superspace) over which $\Psi[q_{ij}]$ is defined.

A distinct canonical quantization program, Loop Quantum Gravity (LQG) \cite{Ashtekar:1981sf,Rovelli:1997yv,Thiemann:2002nj}, also starts from the Hamiltonian formulation but employs a different set of fundamental canonical variables: Ashtekar variables (an SU(2) connection and its conjugate densitized triad). This choice facilitates a non-perturbative, background-independent quantization. While LQG also faces constraint equations analogous to the WDW equation, they are formulated in terms of these new variables, and the Hilbert spaces are constructed differently, based on spin networks. Both geometrodynamics and LQG, despite their differences, adapt the Hamiltonian viewpoint to tackle the quantum dynamics of gravity.

A major challenge common to full canonical quantum gravity, particularly in geometrodynamics, is the infinite-dimensional nature of superspace. Solving the WDW equation, or its LQG counterpart, in such a setting is notoriously difficult. This difficulty motivated researchers to look for tractable truncations, leading to minisuperspace models.
Minisuperspace models reduce the infinite degrees of freedom to a finite (often small) number by imposing symmetry assumptions (e.g., homogeneity, isotropy, or spherical symmetry) on the metric (or connection/triad variables) before quantization~\cite{Hartle:1983ai}. Examples include homogeneous cosmologies, which form the basis of Loop Quantum Cosmology (LQC)~\cite{Bojowald:2008pu,Agullo:2016tjh} (a symmetry-reduced application of LQG principles), as well as models for spherically symmetric black holes~\cite{Kuchar:1994zk,Ashtekar:2005qt,Gambini:2013ooa,Modesto:2008im,Grumiller:2002nm}. In these reduced models, the constraint equations become ordinary (or partial, if some inhomogeneity remains) differential equations, making them more amenable to analytical and numerical treatments. Despite their simplifying assumptions, minisuperspace models often capture essential physical features anticipated from a full theory---such as singularity resolution or modified horizon dynamics---and have played a key role in providing insights into quantum cosmology and black hole quantization.

%%%%%%%%%%%%%%%%%%%%
\subsection{The Problem of Time}

A key conceptual hurdle in canonical quantum gravity is the ``problem of time''~\cite{Anderson:2012vk,Kiefer:2013jqa,Gielen:2020abd,Rovelli:1989jn}. In general relativity, the Hamiltonian is a constraint vanishing on-shell, precluding an explicit ``external'' time parameter; this contrasts with typical quantum theories where a Hamiltonian generates temporal evolution. The resulting tension, emphasized by Anderson and others, emerges when GR constraints are promoted to quantum operators, causing the standard Schr\"odinger-like notion of time evolution to disappear. Attempts to resolve this include redefining time via internal degrees of freedom (e.g., matter fields or geometric variables) or relational interpretations of Wheeler--DeWitt equation solutions. Nevertheless, the problem of time remains a crucial open question in forging a consistent theory of quantum gravity.

%\textcolor{red}{with the help of AI: A major challenge in canonical quantum gravity is the "problem of time"~\cite{Anderson:2012vk,Kiefer:2013jqa,Gielen:2020abd,Rovelli:1989jn}. In general relativity, the Hamiltonian vanishes on-shell, leaving no external time parameter. Quantum mechanics, however, relies on a Hamiltonian-generating evolution, creating tension when GR’s constraints are quantized. Proposed solutions—such as defining time relationally via matter fields or geometric variables, or interpreting the Wheeler–DeWitt equation—have not fully resolved the issue, leaving it central to constructing a consistent quantum gravity theory.}

%%%%%%%%%%%%%%%%%%%
\subsection{Research Idea and Structure of the Paper}
The idea of this paper is to work within the realm of minisuperspace models and explore the quantum structure of static, spherically symmetric spacetimes. However, given the problem of time mentioned above, we do not insist on an ADM-type quantization. To circumvent this type of issues, we can remember the fact that the choice of the quantization hypersurface can strongly affect the constraints and algebraic structure of the corresponding quantum theory. 
This effect and its physical implications has been extensively explored, for example in the context of light-cone quantization of Quantum Chromo Dynamics~\cite{Burkardt:1995ct,Brodsky:1997de,Bakker:2013cea}, or more recently in  QG~\cite{Wieland:2025qgx}.
Even more, it has been argued that a quantization surface should be chosen by the symmetry that best reflects the physical situation of a given problem~\cite{Oeckl:2003vu,Oeckl:2005bv,Oeckl:2022mvg}.
In this spirit,  we exploit the previously imposed spherical symmetry and introduce and explore a Static Equal Radius (SER) quantization prescription, instead of the more frequently used equal time or light-cone quantization.

This is realized by implementing the following steps:
\begin{enumerate}[label=\Alph*]
    \item Choose a theory expressed in terms of a Lagrangian. Restrict to spherically symmetric settings by adopting spherical coordinates $t, r, \theta, \phi$. Then reduce the action to the radial degrees of freedom and fix the gauge if necessary. Write the action as a functional of the remaining degrees of freedom, $S = S(f,h)$, and integrate over the decoupled coordinates such that $S = \int dr\, L(f(r), g(r))$.
    \item Perform a Legendre transformation of the Lagrangian to obtain a Hamiltonian, $L \rightarrow H(p,q)$, treating the non-trivial direction, which is orthogonal to the quantization hypersurface as direction of evolution. This evolution is then described by the afore mentioned Hamiltonian $H(p,q)$.
    For the case of static spherical symmetric surfaces, this direction is the radial variable $r$. Please note that, in order to give these Hamiltonian equations the usual dimensions, we will eventually redefine $r\rightarrow r/c$. Note further, that even though this variable then has units of time and one may call it somewhat misleadingly radial time, it is by no means the ``time'' which is given from the metric signature, which in the classical context even changes sign at the crossing of a horizon. Thus, it is the choice of quantization hypersurface which defines the direction of the Hamiltonian evolution parameter $r$ and not the special entrance in a given classical space time metric.
    \item From here on, one can proceed with the standard machinery of the canonical approach to quantum mechanics, which allows us to derive uncertainty relations for the observables of the theory.
\end{enumerate}

In the following subsection, we revisit some standard notation from canonical quantum mechanics. Then, in Section~\ref{sec_QunatGR}, we apply the same method to General Relativity (GR). The results, in terms of uncertainty relations, are calculated and discussed in Sections~\ref{sec_Uncert} and~\ref{sec_Disc}. Final conclusions are presented in Section~\ref{sec_Concl}.

%%%%%%%%%%%%%%%%%
\subsection{The Free Particle: A Brief Review}
\label{subsec_free}

To establish essential notation and concepts for the subsequent analysis, we concisely review the non-relativistic free particle. The system is described by the action
\be\label{eq_SPP}
S_{PP}= \int dt\, L = \int dt\, m\frac{\dot{x}^2}{2}.
\ee
From this Lagrangian $L = m \dot{x}^2/(2)$, the conjugate momentum $p = \partial L / \partial \dot{x}$ leads to the Hamiltonian $H = p^2/(2m)$. Canonical quantization then replaces classical variables with operators and Poisson brackets with commutators, yielding the Heisenberg equation of motion for an operator $\hat{A}$:
\be\label{eq_eomHeisi}
i \hbar \frac{d}{dt} \hat{A_H} = [\hat{A}_H, \hat{H}] + i \hbar \left( \frac{\partial \hat{A}_S}{\partial t}\right)_H, % Verify i\hbar factor for the partial derivative term
\ee
where the labels $H$ and $S$ indicate the operator in Heisenberg and Schr\"odinger picture respectively. Further details, including the explicit definition of Poisson brackets, classical equations of motion, the derivation of expectation values $\langle \hat{\mathcal{O}} \rangle$ for relevant operators, are textbook material.
One of the results which is relevant for the later discussion is based on the Robertson uncertainty principle, which for any two Hermitian operators $\hat{A}$ and $\hat{B}$ states:
\be\label{eq_UncertQM0}
\sigma_A^2 \sigma_B^2 =
\left(
\langle \hat{A}^2 \rangle - \langle \hat{A} \rangle^2
\right)
\left(
\langle \hat{B}^2 \rangle - \langle \hat{B} \rangle^2
\right)
\ge
\left| \frac{1}{2i} \langle [\hat{A}, \hat{B}] \rangle \right|^2.
\ee
For the particular case of $\hat x$ and $\hat p$ it
shows how quantum uncertainties 
\be\label{eq_UncertQM1}
\sigma_x^2(t) \sigma_{p_x}^2(t) =
(p_{2,0} - p_0^2)
\left[
(x_{2,0} - x_0^2) - t \frac{2}{m} (p_0 x_0 - C_{1,1} )+ \frac{t^2}{m^2} (p_{2,0} - p_0^2)
\right]
\ge
\frac{\hbar^2}{4},
\ee
evolve (grow) over time.
Here, $x_0,\,p_0,\,x_{2,0}, \,p_{2,0}$ and $C_{1,1}$ are the integration constants that arise from integrating the expectation value of (\ref{eq_eomHeisi}) for the operators $\hat x,\, \hat p,\, \hat x^2,\,  \hat p^2, $ and $(\hat x \hat p+ \hat p \hat x)/2$.  For more details regarding this section and in particular the derivation of eq.~\ref{eq_UncertQM1} see Appendix~\ref{subsec_free_appendix}.

With this foundational quantum mechanical framework outlined, we proceed to our objective, the quantization of spherically symmetric spacetimes, which will follow the philosophy summarized in this subsection.

%%%%%%%%%%%%%%%%%%%
\section{Quantization in General Relativity}
\label{sec_QunatGR}
%???Ben is here???

Let's now consider the Einstein Hilbert action with a cosmological constant
\be\label{eq_action0}
S_{GR}=\int d^4x \sqrt{-g}c^3
\frac{R+2\Lambda}{16 \pi G}.
\ee
Now, we apply the steps A-F sketched in the previous section to the action (\ref{eq_action0}).
%%%%%%
\subsection{Coordinates}

Imposing static spherical symmetry
the line element reads
\be\label{eq_LineElement}
ds^2= n^2(r) g(r) c^2d t^2-\frac{1}{g(r)} dr^2-r^2d\theta^2-r^2\sin^2(\theta)d\phi^2.
\ee
This line element only has two unknown functions $n(r)$ and $g(r)$.

%%%%%%
\subsection{Reduced Action}\label{sec:ReducedAction}

Inserting the metric (\ref{eq_LineElement}) into the action (\ref{eq_action0}),  integrating over $\int d\theta d\phi$, and factorizing the time integral $\int dt=Z$ ( this could be understood as an arbitrary finite interval of time)
we are left with
\bea\label{eq_action1}
S&=&\int dr L(n(r),g(r))\\
&=&\int -\frac{c^4 Zdr}{4 G }\left(
r^2 \dot g \dot n + n(-2-2r^2\Lambda +2 g + 4 r \dot g + r^2 \ddot g)
\right),
\eea
where we used the change in notation
$\partial_rg \rightarrow \dot g$.
The Lagrangian in (\ref{eq_action1}) can be more conveniently written by integrating $\ddot{g}$ by parts. We get
\be\label{eq_action2}
S=B-\int dr \;
n(r)
\frac{c^4 Z}{2 G}\left(-1-r^2 \Lambda + g + r \dot g\right),
\ee
where the boundary term reads
\be
B=\left.\frac{c^{4} r^{2} Z}{4G_{0}\,|n|}\,\frac{d}{dr}\!\left(n^{2}g\right)\right|^{r_f}_{r_i}.
\ee
In what follows, we will use and discuss properties in the bulk of the theory $r_i<r<r_f$, where the contribution of the boundary term is just a constant, which can be factored out. However, for global properties, such as entropy or global Bondi mass, one has to consider the non-trivial contribution of $B$.

%%%%%%
\subsection{Legendre Transformation and Poisson Brackets}

When passing to a Hamiltonian formulation, we want to make sure that the remaining variable has the convenient unit of time. For this purpose we introduce the change of variables $\tilde r =r/c$. After this change, we drop the tilde notation, but we remember that from here on (or more precisely from now on), the variable denoted as $r$ is measured in seconds.
Now, we obtain the dimensionless canonical momenta for the fields $n$ and $g$ from (\ref{eq_action2})
\bea\label{eq_canmom}
p_n&=&\frac{\partial L}{\partial\dot n}=0,\\
p&=&\frac{\partial L}{\partial\dot g}=n(r)\frac{c^5 Z r}{2G},\label{eq_poff}
\eea
With these momenta, we can write the canonicial Hamiltonian as
\bea\label{eq_Ham1}
H_c&=& p_n \dot n + p \dot g- L\\ \nonumber
&=&\frac{(1+r^2 (-\Lambda) c^2 -g)p}{r}.
\eea
This Hamiltonian is not unique.  It has to be extended by the two relations (\ref{eq_canmom} and \ref{eq_poff}), which can be implemented
in terms of constraints.
We choose
\bea\label{eq_Constraint1}
\phi_1&=& p_n,\\
\phi_2&=& \left(
\frac{p}{r}-n \Gamma
\right),\label{eq_Constraint2}
\eea
where we defined the constant
\be\label{eq_GammaDef}
\Gamma = \frac{c^5 Z}{2 G},
\ee
which has units of energy.

With these constraints, the full Hamiltonian reads
\be
H=H_c+ \lambda_1 \phi_1 + \lambda_2 \phi_2.
\ee
The Poisson brackets for this system are
\be
\{A,B\}\equiv\frac{\partial A}{\partial g}\frac{\partial B}{\partial p}-
\frac{\partial B}{\partial g}\frac{\partial A}{\partial p}+
\frac{\partial A}{\partial n}\frac{\partial B}{\partial p_n}-
\frac{\partial B}{\partial n}\frac{\partial A}{\partial p_n}.
\ee
Let's first evaluate the canonical equations of motion for the constraints and their Lagrange multipliers
\bea
\phi_1&=&0,\\
\phi_2&=&0,\\
\frac{d\phi_1}{dr}&=& \{\phi_1,H \}+ \frac{\partial \phi_1}{\partial r}=r \Gamma \lambda_2
\\
\frac{d\phi_2}{dr}&=&\{\phi_2,H \}+ \frac{\partial \phi_2}{\partial r}=
p-r\Gamma (n+r \lambda_1).
\eea
The form of the constraints can not depend on $r$. Since this requirement can be fulfilled by the Lagrange multipliers, they are second class.
We solve this condition by imposing
\bea\label{eq_l2sol}
\lambda_2&=&0\\
\lambda_1&=&\frac{p-n r \Gamma}{r^2 \Gamma}.\label{eq_l1sol}
\eea
Now we derive the remaining equations of motion, replace the Lagrange multipliers (\ref{eq_l2sol}, \ref{eq_l1sol}) and solve 
%\bea
%\dot p &=& \frac{p}{r}\\
%\dot g &=&\frac{1-g+r^2 (-\Lambda) c^2}{r}+\lambda_2 \\
%\dot n &=&0\\
%\dot p_n &=&r \Gamma \lambda_2.
%\eea
%In these equations, we can now 
%giving
%\bea
%\dot p &=& \frac{p}{r}\\
%\dot g &=&\frac{1-g+r^2 (-\Lambda) c^2}{r}\\
%\dot n &=&0\\
%\dot p_n &=&0.
%\eea
%The solutions for 
%
$p,\,p_n,$ and $g,\, n$ 
\bea
p &=&C_{0,1} r\\
g &=& 1 + \frac{C_{1,0}}{r}+\frac{1}{3}r^2 (-\Lambda) c^2\\
  p_n &=&p_{n,0}\\
  n &=& n_0.
\eea
We set the integration constant
\be
C_{1,0}=-\frac{2 G M_{0}}{c^{2}}
\ee
so that the metric function \(g\) reproduces the classical
Schwarzschild form.  
In classical GR the parameter \(M_{0}\) is itself an integration
constant, fixed by demanding that geodesic motion reduces to Newton’s
law in the weak-field limit.

Compatibility with the Schwarzschild time–time component \(g_{00}\)
further requires
\be
n_{0}=1=\frac{p}{\Gamma r}=\frac{C_{0,1}}{\Gamma}.
\ee
With these identifications the classical field equations are solved
consistently.

Note that instead of fixing the constants $n_0$ and $C_{0,1}$, we also could have rescaled the time variable of our theory.

%%%%%%%%%%%%%
\subsection{Dirac Brackets}
As discussed in Ref.~\cite{Dirac:1950pj,dirac2013lectures,Date:2010xr}, to prepare quantization we first have to derive the matrix of the Poisson brackets of the constraints
\be
\{\phi_i,\phi_j \}=P_{ij}.
\ee
We find that
\bea
P_{ij}=\Gamma^2 r^d \left( 
\begin{array}{cc}
     0&1  \\
    -1 & 0
\end{array}
\right)&,\quad &P_{ij}^{-1}=\frac{1}{r^d 
 \Gamma^2} \left( 
\begin{array}{cc}
     0&-1  \\
    1 & 0
\end{array}
\right).
\eea
To make sure that operators do commute with the constraints $\phi_i$, we define the Dirac brackets
\be\label{eq_DiracBrack}
\{ A,B\}^*= \{A,B \}-\{A,\phi_i \}P_{ij}^{-1}\{B,\phi_j \}.
\ee
It is straight forward to check, that any
operator commutes with the constraints $\phi_i$. Thus, since the constraints (\ref{eq_Constraint1} and \ref{eq_Constraint2}) are second class, they become strongly zero, which we can use as external identities and not just as results of the equations of motion
\bea\label{eq_Constraint1S}
 p_n &=&0,\\
n &=& \frac{p}{r \Gamma}\label{eq_Constraint2S}.
\eea

%%%%%%
\subsection{Quantization}
The next step consists of quantizing the system. We follow the discussion on quantization of Hamiltonian constraint systems with second class constraints in Ref.~\cite{Dirac:1950pj,dirac2013lectures,Date:2010xr} and impose the usual canonical relations for the Dirac brackets~(\ref{eq_DiracBrack}) and the variables $g$ and $p$
\bea
\{ A,B\}^* &\rightarrow & [\hat A,\hat B]\\ \label{eq_ghat_H}
g &\rightarrow & \hat g(r)=g(r),\\ \label{eq_phat_H}
p&\rightarrow & \hat p = - i \hbar \frac{\partial}{\partial g(r)},\\ \label{eq_HQ}
H&\rightarrow &\hat H=\frac{1-r^2\Lambda c^2}{r}\hat p-\frac{(\hat g \hat p)_W}{r}.
\eea
It should be mentioned that the above operators are in Heisenberg picture. The operators in the Schr\"odinger picture are independent of the variable r. The commutation relation $[\hat g, \hat p]=i\hbar$ follows directly.
We note that products of operators, such as the $\hat g \hat p$ appearing in the Hamiltonian, are symmetrized by Weyl ordering in the field and momentum variable $\hat g \hat p\rightarrow (\hat g \hat p)_W$, to make it hermitian. 
The Weyl ordering prescription applied to an operator of the form $(\hat g^n \hat p^m)_W$ contains all different permutations of the operators, multiplied by $n! m!/((n+m)!)$. 
Different ordering could potentially lead to ordering ambiguities. 
In the following, we will stick to Weyl ordering for all operators.
For a brief discussion of what motivated our choice of ordering prescription, see Subsection~\ref{subsec_Weyl}.
The Heisenberg equations allow to calculate the $r$ evolution of the expectation value of any quantum operator $\hat A$ (\ref{eq_eomHeisi}).
The expectation value of this equation provides us with a generalization of the Ehrenfest theorem 
\be\label{eq_Ehrenfest}
i \hbar\frac{d \langle \hat A_H \rangle _H }{dr}= +
\langle [\hat A_H,\hat H] \rangle _H + i \hbar  \langle \left( \partial_r \hat A_S \right)_H \rangle _H.
\ee
Note that in the Schr\"odinger picture the operators $\hat g$ and $\hat p$ are r-independent. Hence the last term of (\ref{eq_Ehrenfest}) will vanish for all operators such as e.g. $(\hat g^a \hat p^b)$.
The partial derivative will only act on manifestly $r$-dependent expressions, such as it appears for example in the Hamiltonian~(\ref{eq_HQ}).

Further, we promote the strong identities (\ref{eq_Constraint1S} and \ref{eq_Constraint2S}) to operator identities. This implies, in particular, that the operator of the time-time component of the metric (\ref{eq_LineElement}) reads,
\bea
\hat g_{0,0}&=&(\hat n^2 \hat g)_W
=\frac{( \hat g \hat p^2)_W}{r^2\Gamma^2}\label{eq_Constraint2O}.
\eea
In our work, we set our framework by assuming the existence of a Hilbert space $\mathcal{H}=L^2( \mathbb{R},d\mu)$ with orthonormal set of basis $\{ \phi_n (g)\}$ where our formalism lives. For more details regarding the discussion on the Hilbert space, we refer the reader to the section~\ref{subsec_SchroeHeis} and Appendix~\ref{sec_AppendD}. In the coming sections where we evaluate the expectation values, we use the Ehrenfest theorem given in (\ref{eq_Ehrenfest}). There we calculate the evolution of the expectation values and for such calculation we can assume that the inner product with suitable measure exist. Given our assumption that such expectation values exist, we can calculate them directly using their commutator with  the Hamiltonian and solving the differential equation in (\ref{eq_Ehrenfest})

%%%%%%
\subsection{Expectation Values}

Applying (\ref{eq_Ehrenfest}) to quantum operators, we have to make sure that the operators are properly defined.

We obtain
\bea\label{eq_1vev}
\langle 1 \rangle&=& C_{0,0}\\ \label{eq_gvev}
\langle \hat g \rangle&=&C_{0,0}+\frac{C_{1,0}}{r}+C_{0,0}\frac{r^2 (-\Lambda) c^2}{3}\\ \label{eq_pvev}
\langle \hat p \rangle&=& C_{0,1} r\\ \label{eq_gpvev}
\langle 
(\hat g \hat p)_W\rangle &=&
r C_{0,1}+\frac{1}{3}r^3 (-\Lambda) c^2 C_{0,1}+C_{1,1}\\ \label{eq_p2vev}
\langle \hat p^2\rangle&=& r^2 C_{0,2}\\ 
\label{eq_g2vev}
\langle \hat g^2\rangle&=& C_{0,0}+\frac{2C_{1,0}}{r}+\frac{C_{2,0}}{r^2}+\frac{2 C_{1,0} r (-\Lambda) c^2}{3}+C_{0,0}\frac{2 r^2 (-\Lambda) c^2}{3}+C_{0,0}\frac{r^4 (-\Lambda)^2 c^4}{9}\\
\label{eq_pgpvev}
\langle ( \hat g \hat p^2)_W\rangle&=& r C_{1,2}+r^2 C_{0,2}+ \frac{1}{3}r^4 (-\Lambda) c^2 C_{0,2}
\\
\langle (\hat g^2 \hat p^2)_W\rangle&=&
C_{2,2}+ 2 r C_{1,2}+r^2 C_{0,2}+\frac{2}{3}r^3 (-\Lambda) c^2 C_{1,2}+\frac{2}{3}r^4(-\Lambda) c^2 C_{0,2}+ \frac{1}{9} r^6 (-\Lambda) c^2 C_{0,2}
\label{eq_pgpgvev}
\eea
where $C_{a,b}$ are the corresponding real valued integration constants. Each expectation value is obtained by integrating a first order differential equation, thus each expectation value contributes one new integration constant to the system.
The index $a$ of these constants stands for the power of $\hat g$ operators, while the index $b$ stands for the power of $\hat p$ operators.
We realize that the expectation values of any of these operators can be written as
\be\label{eq_VevGeneral}
\langle (\hat g^a \hat p^b)_W\rangle=
\sum_{i,j,k\ge 0}^{i+j+k=a} r^{b-j+2k}
C_{j,b}
\left(\begin{array}{ccc}
    & a&  \\
     i&j&k 
\end{array}\right)
\left( \frac{(-\Lambda) c^2}{3}\right)^k 
\;\;\;\text{with}\;\;\; \begin{pmatrix}
a \\ i\, j\, k
\end{pmatrix}
= \frac{a!}{i! \, j! \, k!}.
\ee
Equation~\ref{eq_VevGeneral} can be confirmed by the use of computer algebra programs such as {{\it Mathematica}} by explicitly calculating (\ref{eq_Ehrenfest}) for an arbitrary product of operators $\hat g$ and $\hat p$ with any power $a$ and $b$, and comparing the result with (\ref{eq_VevGeneral}). We have explicitly checked this in {{\it Mathematica}} for higher power of $a$ and $b$. However, for the scope of our work and further analyses, we only work with the VEVs explicitly given above which were all calculated using (\ref{eq_Ehrenfest}) directly.
%%%%%%%%%%%%%%%%%%%
\subsection{The Spacetime State}
The constants \(C_{j,b}\) in \eqref{eq_VevGeneral} parametrize the
spherically symmetric quantum state of spacetime.
The first relation, \eqref{eq_1vev}, enforces normalization of the
associated wave function \(|\psi\rangle\).
We might fix this normalization to unity,
\be\label{eq_C00}
C_{0,0}=1 ,
\ee
like in quantum mechanics, but in the context of a spacetime state this condition might as well be neglected and the constant $C_{0,0}$ retained explicitly.
The remaining constants encode additional features of the spacetime
state $\Psi$. For example, we can expect that the symmetry of $\Psi$ determines many properties and relations among these constants, just as they do in the example we gave for non-relativistic quantum mechanics (\ref{eq_symmetryExample}). It remains to be seen how far this analogy holds, but for now we must be content with the following statement:\\
The numerical value of the integration constants $C_{i,j}$ depends on the quantum state under consideration.
Thus, information on
these constants has to be obtained by exploring the spacetime state with measurements and observations.

%%%%%%%%%%%%%%%%%%%
\subsection{Observables}
\label{subsec_Obs}

In quantum mechanics, an observable is a physical quantity associated with a Hermitian operator. Measurement outcomes correspond to the eigenvalues of this operator, and the expectation value represents the average result of many measurements on identically prepared systems. However, this definition does not make any reference, whether we can or can-not perform the actual observation. Realizing that a clean definition of an observation is an open question in the foundations of quantum mechanics, it is even more subject to discussion in the context of quantum gravity~\cite{Rovelli:1990ph}.

In this subsection we will 
mention three candidates that could serve as ``observable quantities''.

\begin{itemize}
    \item {\bf{Comparison of circumferences}}\\
%    ???I think this is more suble, because of the inverse operator $1/\hat g dr^2$ appearing in the line element.???
    
The expectation value of the operator $\hat g^{-1}$ is directly linked to an
observable radial quantity.  To see this, set $dt=0$ in the line
element~\eqref{eq_LineElement}, so that radial spatial intervals satisfy
$ds^{2} = \hat g^{-1}\,dr^{2}$.  Imagine two neighbouring coordinate
circles far outside the black hole horizon, with circumferences
$L_{1}=2\pi R_{1}$ and $L_{2}=2\pi(R_{1}+\delta r)$.  By construction
$
\delta r \;=\; \frac{L_{2}-L_{1}}{2\pi}.
$
The actual proper distance between the two circles is
$\,\delta s = \sqrt{\langle\hat g^{-1}\rangle}\,\delta r$.  Hence a
measurement of $L_{1}$, $L_{2}$, and $\delta s$ yields
\begin{equation}\label{eq_ginvvev}
\bigl\langle\hat g^{-1}\bigr\rangle(R_{1})
   \;=\;
   \left(\frac{2\pi\,\delta s}{L_{2}-L_{1}}\right)^{2}.
\end{equation}
In practice, evaluating the left–hand side of
\eqref{eq_ginvvev} is non-trivial because of the inverse operator,
as discussed in Appendix~\ref{sec_AppendB}.  By contrast, the
expectation value of~$\hat g$ itself is far easier to obtain; it keeps
the same functional form as the classical metric coefficient:
$
\bigl\langle\hat g\bigr\rangle
   = 1 - \frac{2GM_{0}}{c^{2}r}\bigl(1+\epsilon_{1,0}\bigr)
     - \frac{r^{2}(\Lambda)c^{2}}{3},
$
where we have adopted the normalisation~\eqref{eq_C00} and defined
\begin{equation}\label{eq_C10}
C_{1,0} \;=\; -\frac{2GM_{0}}{c^{2}}\bigl(1+\epsilon_{1,0}\bigr).
\end{equation}
Here $M_{0}$ is the classical mass parameter, while
$\epsilon_{1,0}$ quantifies any quantum deviation from the classical
integration constant.  In the limit $\epsilon_{1,0}\!\to 0$ the quantum
result smoothly reproduces the classical metric coefficient, at least
for the directly measurable quantity $\langle\hat g\rangle$.

\item {\bf{Gravitational redshift}}\\
The gravitational redshift is associated to the metric function $g_{00}$.
To simplify the discussion on redshift, let's assume shortly that the cosmological constant is negligible. 
In this case, we can use the time-rescaling symmetry of the initial Lagrangian such that $\lim_{r \rightarrow \infty} g_{00}=1$.
In this case, the comparison between the local time $d\tau_R$ of a clock at radial coordinate $R$ and the time lapse of another clock at radial infinity $d\tau_\infty$, allows a straightforward definition of a measured gravitational redshift $z$.
This redshift is then related to the metric function $f$ via
\be
g_{00}=\frac{1}{(1+z)^2}.
\ee
The quantum operator for this observable was defined in (\ref{eq_Constraint2O}). By using (\ref{eq_VevGeneral}) we can write its expectation value as
\be\label{eq_g00Vev}
\langle \hat g_{00}\rangle=
\frac{ C_{0,2}}{\Gamma^2}+\frac{ C_{1,2}}{r \Gamma^2}+\frac{r^2 \Lambda c^2 C_{0,2}}{3 \Gamma^2}.
\ee
Thus, the functional form of the classical result is recovered for
\bea\label{eq_C01}
C_{0,2}&=& \Gamma^2(1+\epsilon_{0,2}),\\
C_{1,2}&=&-\frac{2 G M_0 \Gamma^2}{c^3}(1+\epsilon_{1,2}).
\eea
Like before, the parameters $\epsilon_{0,2}$ and $ \epsilon_{1,2}$ account for a possible discrepancy between our classical expectation and the quantum observable. Yet again in the limit $\epsilon_{i,j}\rightarrow 0$ such a discrepancy would vanish.
Note that it is not true that (\ref{eq_g00Vev}) is identical to the product of the corresponding expectation values
$(\langle (\hat g \hat p^2)_W\rangle\neq \langle \hat g\rangle \langle \hat p\rangle \langle \hat p\rangle)$, even though some special wave function with the corresponding integration constants might have this property.

%%%%%%%%%%%%%%%%%%%%%%
\item {\bf{Geodesic motion}}\\
Most of our knowledge about the Universe is obtained by using the concept of geodesics in one way or the other. Thus, it is natural that QG effects on geodesics have been explored from numerous different angles \cite{Dalvit:1997yc,Dalvit:1999wd,Parikh:2020kfh,Chawla:2021lop,Bak:2022oyn,Bak:2023wwo,Cho:2021gvg,Hsiang:2024qou,Piazza:2025uxm,Nitti:2024iyj,Piazza:2022amf,Piazza:2021ojr,Pipa:2018bui}. The geodesic equation contains combinations of products and derivatives of metric functions. In the context of a quantum background, these metric functions naturally become operators 
($g_{\mu \nu}\rightarrow \hat g_{\mu \nu}$ and $\Gamma^\mu_{\alpha \beta} \rightarrow \hat \Gamma^\mu_{\alpha \beta}$)
whose expectation values determine the motion of a test particle.
Let's consider two metric operators $\hat A$ and
$\hat B$. In a quantum theory, the expectation value of the product of these operators is generally not identical to the product of expectation values $\langle (\hat A \hat B)_W\rangle\neq \langle \hat A\rangle \langle \hat B\rangle )$.
Thus, we need to rethink from scratch what it actually means that a particle travels along a geodesic of such a background~\cite{Koch:2023dfz}. We have addressed 
 this question in a separate publication~\cite{Koch:2025qzv}.
%??? don't forget to cite galaxy roation curves https://arxiv.org/pdf/astro-ph/9602099
\item {\bf{Conserved quantities}}\\
It is natural to associate observables to conserved quantities.
From quantum mechanics, as described in subsection~\ref{subsec_free}, the first conserved quantity arises from the Hamiltonian and the conserved quantity is the energy. In contrast in the SER quantization the Hamiltonian (\ref{eq_HQ})  itself depends on the variable $r$ and so does the corresponding eigenvalue
\be\label{eq_HamEV}
\langle \hat H \rangle=
-\frac{C_{1,1}}{r}-\frac{2 C_{0,1}}{3} r^2 \Lambda ^2.
\ee
Thus, (\ref{eq_HamEV}) is not a good observable in this sense (unless $C_{1,1}=C_{0,1}=0$). 
However, it is straight forward to build conserved operators by suitably combining $\hat g$ and $\hat p$ with the radial variable $r$. The simplest example for such an operator would probably be $\hat p/r$, since
\be\label{eq_ConservedP}
\left \langle \frac{\hat p}{r}\right\rangle= C_{0,1}.
\ee
The physical meaning of such conserved quantities and their relation to the no-hair theorem~\cite{Israel:1967za,Heusler:1996ft} is yet to be explored. 
\end{itemize}
We have suggested a number of quantities, which can in principle be used to determine some of the integration constants $C_{i,j}$. However, there are $i\cdot j$ such constants, and thus overwhelmingly more constants than observables that we suggested. Can't we know all constants of a given static spherically symmetric quantum spacetime?
This question can be contemplated from complementary perspectives. On the one hand, there are certainly more observables possible than the few that we just mentioned. On the other hand, 
knowing all constants is forbidden no matter how creative we are in coming up with new observables.
As we will see in the following section, these restrictions arise from the laws of quantum mechanics as a consequence of the uncertainty principle.

%%%%%%%%%%%%%%%%%%%
\section{Uncertainty Relations}
\label{sec_Uncert}

We have seen that in certain limits, our results reproduce the predictions of classical general relativity, by calculating expectation values of the corresponding quantum operators and identifying the integration constants accordingly. 
Now, let's go beyond this and explore the effects that can not be provided by a classical theory.
For this purpose,
let's now revisit the concept of uncertainty relations (\ref{eq_UncertQM0}) for our gravitational theory.

%%%%%%%%%%%%%%
\subsection{[g, p]}

The simplest non-commuting operators are $\hat g$ and $\hat p$. Even though, $\hat p$
is not directly linked to one of the observables mentioned in the previous subsection, it is instructive to explore this scenario. Evaluating (\ref{eq_UncertQM0}) 
for the operators $\hat g$ and $\hat p$
we find 
\be\label{eq_uncertGR00}
\sigma_g^2 \sigma_p^2=
\frac{1}{9}(C_{0,1}^2-C_{0,2})\left(r^2( r^2 \Lambda c^2-3)^2 (C_{0,0}-1)C_{0,0}+6 r (r^2\Lambda c^2-3 )
(C_{0,0}-1)C_{1,0}+9
C_{1,0}^2-9C_{2,0}
\right)
\ge
\frac{\hbar^2}{4}.
\ee
Now, after we impose the normalization condition 
(\ref{eq_C00}),
we realize that all $r$-dependence cancels out and the uncertainty relation reads
\be\label{eq_uncertGR0}
(C_{0,2}-C_{0,1}^2)(C_{2,0}-C_{1,0}^2)
\ge
\frac{\hbar^2}{4}.
\ee
At the moment we have not yet explored observables, that would allow for a straight forward physical interpretation of the integration constants in (\ref{eq_uncertGR0}). We provide an attempt for such an interpretation in these constants in the appendix D.
It is however important to notice that
the cancellation of the $r$-dependence in (\ref{eq_uncertGR0}) is a remarkable feature. It means for example, that the uncertainty between $\hat g$ and $\hat p$ remains well-defined, even at the black hole horizon.

%%%%%%%%%%%%%%
\subsection{[(pgp), g]}

Let's now turn to the uncertainty between the circumference operator, $\hat g$ and the non-commuting part of the redshift operator, which consists of  
 $(\hat g \hat p^2)_W$.
Classically, there is no reason, why the observables $g_{11}$ and $g_{00}$ should not be measured with arbitrary precision. However, in quantum theory
both operators do not commute.
When we evaluate
(\ref{eq_UncertQM0}) for this pair of operators, we get a hardly readable lengthy expression consisting of a numerator and a denominator, which both are functions of $r$.
The inequality, must hold for any value of $r$ and any admittable value of the integration constants.
The system could violate the uncertainty relation if either the numerator gets very small, or the denominator grows stronger than the numerator at $r \rightarrow \infty$.
It is straight forward to realize that the latter case does not occur. 
However, the former case yields a remarkable consequence.
For the sake of simplicity and on the expense of generality let's now consider a quantum state, where most of the new integration constants vanish $C_{i,4}=C_{2,4}=C_{2,0}=C_{1,1}=0$. 
Please note that by considering these constants to vanish, we are implicitly assuming a particularly symmetric spacetime state (which we imagine like assuming that many higher multi-pole moments vanish for a charge distribution in electrostatics). 
Thus, all the findings that result from such a choice are only proven to be true for this class of states.
In any case, for this class of states, the $(gp^2)_W, g$ uncertainty relation can be written as
\be\label{eq_uncertfg0}
\frac{4 G^2 M_0^2 \Gamma^4}{c^2 r^2 C_{0,1}^2}\frac{(c^3 r (3+r^2 (-\Lambda) c^2)-6 G M_0)^2}{3+r^2 (-\Lambda) c^2}\ge \hbar^2.
\ee
We explore this relation in two regimes:
\begin{itemize}
\item {\bf{Small radius}}\\
When one plots (\ref{eq_uncertfg0}) one realizes, that this relation is well-defined, even in the vicinity of the Schwarzschild horizon. This feature can be explored 
in the small radius regime $r^2\ll 1/\Lambda$, where the numerator of (\ref{eq_uncertfg0}) has a minimum at
\be
r_{Mi}=\frac{2 G M_0}{c^3},
\ee
which is exactly the Schwarzschild radius for the classical solution. At this minimum, the inequality (\ref{eq_uncertfg0}) reads
\be\label{eq_uncertfgSS}
\frac{64 G^5 \hbar M_0^6 ( 2 \pi \gamma)^4 (\Lambda)^2 c^3}{ (2 \pi \tilde C_{0,1})^2(3 c^6 + 4 G^2 M_0^2 (-\Lambda) c^2)^2}\ge \hbar^2,
\ee
where we wrote two constants in terms of dimensionless parameters $\Gamma=2 \pi \gamma (M_p c^2)$ and $C_{0,1}=2 \pi \tilde C_{0,1}M_p c^2$, by introducing the Planck mass $M_p^2=\hbar c /G$.
If we read the left hand side of (\ref{eq_uncertfgSS}) as a function of $(\Lambda)$,
we realize that this function has a single minimum at $\Lambda=0$.
Thus, (\ref{eq_uncertfgSS}) provides a lower bound on the absolute value of the cosmological constant, or equivalently 
a lower bound on the central mass in a Universe with a given cosmological constant
\be\label{eq_LambdaLower}
M_0^6\ge \frac{9 (2 \pi \tilde C_{0,1})^2}{64(2 \pi \gamma)^4}\frac{ c^{9} \hbar}{G^5 \Lambda^2}.
\ee
Even though, we do not know the values of the dimensionless constants $\tilde C_{0,1}$ and $\gamma$, this relation suggests that our quantum description is only consistent if the cosmological constant is not zero.
\item {\bf{Large radius}}\\
In the regime of large radii, the numerator of (\ref{eq_uncertfg0}) only has a minimum if $\Lambda >0$. The minimum is reached at the outer radius
\be
r_{Mo}= \sqrt{\frac{6}{(\Lambda) c^2}},
\ee
which is not the radius of the outer horizon of the classical solution.
At this outer minimum, the uncertainty (\ref{eq_uncertfg0}) reads
\be\label{eq_uncertfgCH}
\frac{2(2\pi)^2 G^2 M_0^3 \gamma^4}{3 c^6 \tilde C_{0,1}^2}\left(\sqrt{6} c^2-2 G M_0 \sqrt{(\Lambda)}
\right)^2\ge \hbar^2.
\ee
Again, we do not know the numerical value and physical interpretation of the constants. Nevertheless, we know as minimal condition that
the parenthesis on the left hand side of (\ref{eq_uncertfgCH}) can not be allowed to vanish.
Further assuming that $M_0>0$, that the range of masses is connected, and using that the other constants outside the brackets are real, provides an upper limit on the cosmological constant, or equivalently an upper bound on the central mass in a Universe with cosmological constant
\be\label{eq_LambdaUpper}
M_0^2< \frac{3 c^4}{2 G^2 |(\Lambda)|}.
\ee
Note that from a mathematical point of view, the above inequality could also be in the opposite direction, but this would not make sense from an observational point of view since we already know of many black hole masses that obey the inequality (\ref{eq_LambdaUpper}) and not the inverted inequality.
\end{itemize}
We will discuss the implications of the lower (\ref{eq_LambdaLower}) and the upper bound (\ref{eq_LambdaUpper}) in section \ref{sec_Disc}.

%%%%%%%%%%%%%%
\subsection{Further Uncertainties}

Following the same steps as in the previous subsections, we can calculate the uncertainty relations for all sorts of operators like
$\hat H, \hat p^2, \dots$.
Using the operator identities it is straight forward to see that most of these operators 
do not commute with each other.
Thus, with sufficient patience one can proceed
to use (\ref{eq_VevGeneral})
and calculate the corresponding uncertainty relations. However, since such higher order uncertainty relations involve higher order integration constants, the result can not be  interpreted easily, since we lack physical intuition on these constants. 

Another intriguing issue is the mutual interplay between the
gravitational uncertainty relation,
\(\sigma_{g}^{2}\sigma_{p}^{2}\ge\hbar^{2}/4\)
from~\eqref{eq_uncertGR0}, and the usual quantum–mechanical
uncertainty, \(\sigma_{x}^{2}\sigma_{p_x}^{2}\ge\hbar^{2}/4\)
from~\eqref{eq_UncertQM1}.  
Two complementary limits can be distinguished.

\begin{itemize}
\item \textbf{Geometry-dominated regime.}  
      Geodesic measurements (Subsection~\ref{subsec_Obs}) probe spacetime uncertainties, such as 
      the pair \((\sigma_{g},\sigma_{p})\) that characterises the
      quantum spacetime.
      The test particle’s intrinsic spread
      \((\sigma_{x},\sigma_{p_x})\) then acts as a small perturbation and we can write schematically
      \be
        \sigma_{g}^{2}\,\sigma_{p}^{2}\;\ge\;
        \frac{\hbar^{2}}{4}
        \Bigl(1+
          \frac{\sigma_{x}^{2}\,\sigma_{p_x}^{2}}%
               {S_{\mathrm{GR}}^{2}}\Bigr),
      \ee
      where \(S_{\mathrm{GR}}\) is the gravitational action
      defined in~\eqref{eq_action0}.  Because
      \(S_{\mathrm{GR}}/\hbar\gg1\) for a geometry dominated regime, the
      correction term is suppressed.

\item \textbf{Particle-dominated regime.}  
      When the particle’s Heisenberg spread is the primary quantity,
      the metric fluctuations provide a subleading correction. Now we can write schematically:
      \be
        \sigma_{x}^{2}\,\sigma_{p_x}^{2}\;\ge\;
        \frac{\hbar^{2}}{4}
        \Bigl(1+
          \frac{\sigma_{g}^{2}\,\sigma_{p}^{2}}%
               {S_{\mathrm{PP}}^{2}}\Bigr),
      \ee
      with \(S_{\mathrm{PP}}\) being the world-line action introduced
      in~\eqref{eq_SPP}. 
      As long as we are in a particle-dominated regime, the gravitational correction in the spacetime region under consideration is suppressed.
\end{itemize}
Each inequality thus corrects the other in the spirit of a
generalised uncertainty principle; cf.\
\cite{Hossenfelder:2012jw,Adler:1999bu,Kempf:1994su,Ali:2009zq}.

A serious study of the tasks that were only sketched in this subsection is postponed to future work.

%%%%%%%%%%%
\section{Discussion}
\label{sec_Disc}

Now, we would like to touch on 
different aspects of this work, which have not been elaborated so far:

%%%%%%%%%%%%%%%%%%%%%%%%%%%%%%%
\subsection{Large Scale Scale-dependence}

Traditional renormalisation schemes for quantum gravity on a flat
background—e.g.\ perturbation theory with a momentum cut-off—suggest
that scale dependence sets in only above the Planck mass,
$M_{\rm Pl}\!\simeq\!10^{19}\,\text{GeV}$~\cite{Donoghue:2012zc}.
Such a conclusion is of limited phenomenological use and in any case
fails for observables that vanish in classical GR, where even tiny
quantum corrections are relevant~\cite{Reyes:2023fde,Reyes:2023ags,Koch:2022cta}.
More importantly, perturbation theory itself is inadequate for gravity,
so scale-dependent effects need not be confined to trans-Planckian
energies.

From a renormalisation-group perspective, Wetterich has shown that the
argument of~\cite{Donoghue:2012zc} collapses once a positive
cosmological constant is included: infrared instabilities in the
graviton propagator invalidate perturbation
theory~\cite{Wetterich:2018qsl}.  Reuter’s functional RG programme
likewise demonstrates that gravity is highly sensitive to such IR
dynamics, motivating extensive investigations of their large-scale
implications~\cite{Reuter:1996cp,Reuter:2001ag,Reuter:2004nx,Reuter:2008wj,Nagy:2013hka,Christiansen:2012rx,Christiansen:2014raa,Christiansen:2015rva,Biemans:2016rvp,Denz:2016qks,Biemans:2017zca,Bertini:2024onw,Wetterich:1994bg,Bonanno:2001hi,Bentivegna:2003rr,Koch:2010nn,Canales:2018tbn}.
Similar large-scale effects have also been predicted in the context of Rindler forces, potentially arising from QG~\cite{Grumiller:2011gg}.

The fact that the inequalities (\ref{eq_LambdaLower}) and (\ref{eq_LambdaUpper})
are relevant at enormous mass- and distance scales, leads us to a similar overarching conclusion: \emph{quantum
effects of gravity can remain important far beyond the Planck
length}.

%%%%%%%%%%%%%
\subsection{Choice of the Quantization Hypersurface}

In our understanding, the only unconventional step of this work was to impose the SER quantization prescription which works on an equal-radius hypersurface and consequently identifying the corresponding commutator with a radial derivative.
We know how quantization is performed on the solid ground of a flat equal-time surface. However, when it comes to gravity, this ground becomes shaky. 
Usually, one still attempts to define a constant time hypersurface (HS) with a time-like normal vector and quantize on this arbitrarily chosen surface~\cite{Lochan:2022dht}. 
To the contrary, in our above discussion the hypersurface is a sphere, and the normal vector is a space-like vector in radial direction. Once this is accepted, all following steps are literally canonical.

{{\it
We are not sure how drastic, from a conceptual standpoint, the choice of the SER framework really is.
In GR the choice of HSs is arbitrary anyhow, since these depend on coordinate choices and spacetime curvature.
What we can note here is that our choice does not lead to any direct inconsistencies.}}
As we have pointed out in the introduction, the choice of a quantization hypersurface which represents the symmetry of a given problem has been contemplated in the literature~\cite{Oeckl:2003vu,Oeckl:2005bv,Oeckl:2022mvg}, for example in light-cone quantization
~\cite{Burkardt:1995ct,Brodsky:1997de,Bakker:2013cea,Wieland:2025qgx}.
In this context we want to mention
Carrollian theories~\cite{Donnay:2022aba,Donnay:2019jiz}, where space is absolute (as opposed to time). Thus, in such theories, one  needs a quantization prescription for space-like HSs anyhow~\cite{Cotler:2024xhb}.
Further, there are gravitational theories, where the role of time and radial coordinates are interchanged, are the Kantowski-Sachs (KS) spacetimes~\cite{NotesKantowskiSachs,Chiou:2008eg,Joe:2014tca,Modesto:2004wm,Collins:1977fg}. The Kantowski--Sachs spacetimes are homogeneous but anisotropic models with spatial topology $\mathbb{R}\!\times\!S^2$. The line element of such spacetimes can be written as 
\begin{equation}
ds^2 = -N^2(t)\,dt^2 + a^2(t)\,dr^2 + b^2(t)\,d\Omega^2.
\end{equation}
They describe the interior region of the Schwarzschild solution and similar metrics, where, as mentioned before, the roles of $t$ and $r$ are interchanged, namely the radial variable becomes timelike, while $t$ becomes spacelike. In Ref.~\cite{Joe:2014tca}, the quantization of such a spacetime has been discussed where the main motivation for quantization is to resolve the singularity problem. Different prescription has been discussed and it was shown that only one prescription was able to resolve the singularity problem in the framework considered (see Ref.~\cite{Joe:2014tca} for more details). However, quantization in such spacetimes is also along the radial direction as in SER quantization prescription. It is worth noting that, in the above references for KS spacetime quantization, the region inside a black hole horizon is being considered while in SER we are considering region outside the black hole horizon. For instance, in Ref.~\cite{Lenzi:2025pmk}  nonrotating neutral black holes within a canonical framework have been studied with a hybrid quantization prescription. There, inspired from the interior solutions, the authors extend their results to the exterior region and adopt a radial evolution as well. Our motivation for the SER quantization prescription was to avoid the problem of time and whether the difficulties in quantization prescription for KS spacetimes could be connected to SER quantization prescription remains to be explored in future studies.

In general, the study of gravitation systems with radial Hamiltonian has recently gained further attention. For instance, in Ref.~\cite{Livine:2025soz}, the authors study the spherically symmetric spacetime in the context of Einstein-Hilbert action and its effective dynamics where the metric functions are functions of radius $r$ and become the canonical variables of the model. In addition in Refs.~\cite{Blacker:2024rje,Callebaut:2025zpc}, the link between radial canonical gravity and $\text{AdS}_3-$gravity has been studied, where in Ref.~\cite{Callebaut:2025zpc}, the evolution parameter of the model is the radius $r$ as in our case.

A more formal understanding of our quantization could be possible within the context of 
compositional quantum field theory~\cite{Oeckl:2022mvg}. This theory is formulated and quantized in terms of 
compositional locality instead of temporal locality. The difference between both types of locality is that temporal locality imposes time-slices, 
while compositional locality allows to slice spacetime into arbitrary finite regions (e.g. spheres).

From all these discussions about different quantization hypersurfaces arises one question:
{\it{Is the choice of a different quantization hypersurface different physics?}}
If two calculational methods, would give different predictions for the same observables, they have to be considered as different theories. However, we do not believe that this is the case, when we discuss the choice of these surfaces in the context of static spherically symmetric spacetimes. 
Instead, we believe that a
clever choice can help to tackle problems, which are much harder to solve otherwise. This is just like
in classical mechanics, where
it is helpful to choose suitable coordinate systems and dynamical variables for solving special problems with special symmetries.

%%%%%%%%%%%%%%%
\subsection{Weyl Ordering, Why?}
\label{subsec_Weyl}

Promoting classical phase–space functions \(f(g,p)\) to operators
requires an ordering prescription.  
Different descriptions can give different quantum operators which in turn imply different matrix elements, different expectation values (e.g. real or complex), and different eigenvalues.
There is no universally ``best'' ordering prescription. Typically a choice is made, depending on mathematical convenience, characteristics of the system under consideration, and its intuitive interpretation.
We employ Weyl (totally
symmetric) ordering for three reasons.  
\begin{itemize}
    \item 
Weyl ordering maps any real classical observable to a Hermitian
operator~\cite{Weyl:1927vd,Moyal:1949sk},
so metric expectation values such as
\(\langle\hat g\rangle\) and \(\langle\hat g^{-1}\rangle\) are
guaranteed to be real.  
\item  Our canonical model contains no
natural split into creation and annihilation operators; hence normal or
anti–normal orderings, rooted in that split
\cite{Wick:1950ee,Peskin:1995ev}, are not appropriate.  
\item The Weyl rule preserves a transparent classical–quantum
correspondence via the Moyal bracket, allowing the Ehrenfest theorem to
recover the Schwarzschild–(A)dS solution and to isolate the quantum
effects and uncertainty relations—presented in this work.
\end{itemize}

Alternative prescriptions, such as normal or anti-normal ordering~\cite{Wick:1950ee,Peskin:1995ev}, the Born-Jordan scheme, or the covariant Laplace-Beltrami choice~\cite{DeWitt:1952js}, are indispensable in perturbative quantum field theory or on curved configuration spaces. However, these use creation and annihilation operators (which we do not have), sacrifice automatic Hermiticity, or depend on a specific choice of vacuum. 

Thus, for our purposes, the Weyl prescription appears to be the most economical and physically transparent option. Having made this natural choice, it goes without saying that future work may extend our approach by exploring the implementation of alternative operator ordering prescriptions.

%%%%%%%%%%%%%
\subsection{A Comparison to LQC Models for Black Holes}

The interior region of the classical solution of a spherically symmetric  black hole can be described by a cosmological Kantowski--Sachs metric.
This classical analogy triggered a substantial progress in the LQC description of the interior of a black hole, spearheaded by Ashtekar, Olmedo, and Singh (AOS)~\cite{Ashtekar:2018lag,Ashtekar:2018cay,Ashtekar:2020ckv}.
In these models with an effective Hamiltonian it was possible to resolve the issue of the radial singularity at the origin and replace it with a finite bridge between a black- and a white hole.
Further, it was shown that for large black hole masses, the model allows to avoid large QG corrections at the horizon.
Even though subsequent work helped to clarify many aspects and consequences of this proposal (for an involuntarily incomplete list of references see~\cite{Gambini:2020nsf,Carballo-Rubio:2019fnb,Bouhmadi-Lopez:2020oia,ElizagaNavascues:2023gow}).

Our findings with the SER approach are complementary to the program pioneered by AOS in several aspects:
\begin{itemize}
    \item Hamiltonian:\\
    The LQC approach must contend with theoretical uncertainties and technical difficulties arising from the ``{\it{highly intricate nature of the Hamiltonian constraint}}''~\cite{ElizagaNavascues:2023gow}. An interesting twist that appears to address the Hamiltonian problem emerges in unimodular quantum gravity, where the Wheeler--DeWitt equation becomes an evolution equation with respect to the so-called unimodular time~\cite{Gielen:2025ovv}. This work also contains an interesting discussion on the distinction between hermitian and self-adjoint operators in the presence of boundary conditions~\cite{Albrecht:2022sdd} e.g. at $r=0$.

The SER quantization is free of the delicate issue of a vanishing Hamiltonian. Nevertheless, theoretical uncertainties, such as potential operator-ordering ambiguities (i.e.\ alternatives to Weyl ordering), remain present in the SER approach as well.
 
    \item Spacetime region:\\
    The LQC approach is constructed for the region inside of the black horizon and this region is also the focus of its remarkable results concerning the singularity resolution.
    Instead, the focus of the SER quantization is the large radius region because it is in principle accessible to observation. In particular, since we include in our discussion the role of a cosmological constant $\Lambda$, we have found potential QG effects at large distances, which have previously gone unnoticed.
    \item Degrees of freedom:\\
    The number of physical degrees of freedom $N_{dof}$ in constrained Hamiltonian systems is~\cite{HenneauxTeitelboimBook}
    \be
    N_{dof} = N_{can} - N_{FCC} - N_{SCC}/2,
    \ee
    where $N_{can}$ is the number of canonical pairs, $N_{FCC}$ the number of first class constraints, and $N_{SCC}$ the number of second class constraints.
    Both approaches have two canonical pairs. In the AOS formulation there is one first class constraint and no second class constraint. In contrast, 
    the dynamical variables in the SER approach are subject to no first class constraint, but to two second class constraints (\ref{eq_Constraint1}, \ref{eq_Constraint2}).
    Thus, both AOS and SER have a single physical degree of freedom $N_{dof}=1$.
    \item Quantum $\rightarrow$ classical transition:\\
    In the LQC approach, the transition from the quantum regime to the classical regime is realized in terms of two control parameters $\delta_b$ and $\delta_C$, which are introduced in the effective Hamiltonian. In the SER approach, the system is always quantum, but for certain spacetime wavefunctions, the expectation values might be indistinguishable from classical results. Here, the transition to the classical regime is encoded in the values of the integration constants $C_{i,j}$ that characterize the nature of spacetime.
\end{itemize}
Thus, the AOS and the SER approach are complementary in their technique and region of interest. There is however also an interesting regime of overlap, since the Kruskal coordinates also apply for the exterior region as well, it is noted that ``{\it{there is now nontrivial dynamics as one evolves from one timelike homogeneous surface to another in the radial direction. While this is somewhat counter-intuitive at first because this evolution is in a spacelike direction, there is nothing unusual about the setup from the Hamiltonian perspective even for full general relativity}}''~\cite{Ashtekar:2018cay}.
This is an encouraging statement, since it is the ``motion'' in the radial direction that is at the heart of the SER approach.

%%%%%%%%%%%%%
\subsection{It's Heisenberg not Schr\"odinger 
and ``What About the Norm and Hilbert space?''}
\label{subsec_SchroeHeis}
The quantization is performed on a Hilbert space of square-integrable wave functions 
$\Psi(g)$ over the configuration variable $g$. 
We adopt the standard representation familiar from canonical quantum mechanics,
\begin{equation}
\mathcal{H} = L^2(\mathbb{R}, d\mu(g))=L^2(\mathbb{R},\mathcal{J}dg) \qquad , \qquad \|\Psi\|^2 = \langle \Psi | \Psi \rangle.
\end{equation}
The Hilbert space $\mathcal{H} = L^2(\mathbb{R}, d\mu(g))$ is separable and thus admits 
a countable orthonormal basis $\{\phi_n(g)\}$ satisfying 
\[
\langle \phi_m | \phi_n \rangle = \int \phi_m^*(g)\,\phi_n(g)\,d\mu(g) = \delta_{mn},
\qquad 
\sum_n |\phi_n\rangle\langle \phi_n| = \mathbb{I}.
\]
In addition, one may use the continuous set of generalized eigenstates 
$\{|g\rangle\}$ of the operator $\hat g$, which satisfy 
$\langle g|g'\rangle = \mathcal{J}(g)^{-1}\delta(g-g')$ and the completeness relation 
$\int |g\rangle\langle g|\,d\mu(g)=\mathbb{I}$. 
In this representation, any state $|\psi\rangle$ admits the expansion 
$|\psi\rangle = \int \psi(g)\,|g\rangle\,d\mu(g)$ with $\psi(g)=\langle g|\psi\rangle$.
As shown in the quantization section~\ref{sec_QunatGR}, the commutator relation between the operators $\hat g$ and $\hat p$ is satisfied as well. In this work, we are considering constrained Hamiltonian systems. In such systems, one defines a kinematical Hilbert space as the space of all square-integrable functions. However, defining a physical Hilbert space, relies on the nature of the constraints. If one deals with first class constraints, these constraints will also be promoted to operator in the quantization prescription. Thus, one imposes the condition that the physical Hilbert space is defined by acting with the constraints on the wave functional such that they vanish, $\hat \chi_a | \Psi(g)\rangle = 0$.
Here $\hat \chi_a $ are the first class constraints being promoted to operator after quantization. Therefore, the Hilbert space with the presence of the first class constraints is written as $\mathcal{H}_{physical}= \{ \; \Psi \in \mathcal{H}_{kin} \quad | \quad \hat \chi_a | \Psi(g)\rangle = 0 \; \}$. \\
Such a condition would be inconsistent when having second class constraints only. When there are second class constraints, one first goes from Poisson brackets to Dirac brackets and thus these constraints become already strongly zero by construction. Thus, the  quantization on the constraint surface is performed after imposing these constraints. Therefore, the kinematical Hilbert space is already the physical Hilbert space.

The Hamiltonian (\ref{eq_HQ}) depends explicitly on the evolution variable $r$.
Under this circumstance, the Heisenberg picture, where all operators are allowed to be $r$-dependent is the most natural choice.
In mathematical physics, it is well established
how to treat a $r$-dependent Hamiltonian,
as long as it can be considered as small
perturbation of an otherwise energy conserving
Hamiltonian. However, in our case
the $r$-dependence can not be treated in this way. Thus, if one would like to translate our results to $r$-independent operators with $r$-dependent wave-functions, as in the Schr\"odinger picture, one might need 
a different mathematical machinery, which we are currently working on as one of the follow-up projects of this work.
Note that in this context that we do not pretend that the radial coordinate $r$ is actually a ``time'', we only needed the statement that one can quantize spherically symmetric systems on a HS with constant $r$.

Having decided to stick with the Heisenberg picture, we still can ask questions like, ``{\it{how does the wave function $|\Psi\rangle$  look like?}}'' and if this question is answered,
``{\it{how is the inner product between two such wave functions $\langle \Psi | \dots  |\Psi\rangle$ actually defined?}}''.
We can not answer these questions at this point, but we can revisit the quantum mechanical point particle system from subsection \ref{subsec_free} for an analogy.
There, we can choose the representation in position space $|\psi\rangle=\psi(x)$
and define the inner product
in terms of an integral over the position.
By doing this, we can find the eigenstates of an operator with the associated eigenvalues (e.g. for the Hamiltonian operator $\hat H \psi(x,E)=E \psi(x,E)$).
The quantization of the  energy $E$ into discrete levels labeled by $E_n$ arises  from imposing that the eigenstates are normalizable $\langle\psi_n|\psi_n\rangle=1$.
Thus, if the analogy point particle and SER goes through, we could understand the spacetime states as functions of $g$, $|\Psi\rangle=\Psi(g)$ and define the inner product as integral over this variable
\be\label{eq_InnerQG}
\langle \Psi | \hat{\mathcal{O}}|\Psi\rangle
=\int_{-\infty}^\infty dg\;{\mathcal{J}} \;  \left(\Psi^*(g) \hat{\mathcal{O}}
\Psi(g)\right)_W
\ee
where ${\mathcal{J}}$ is a properly defined measure. One can show simply that the operators considered in this work satisfy the symmetry and the self-adjoint properties with a measure ${\mathcal{J}}$ being a constant and set to one. We will discuss this in another publication more precisely.
Now, we can speculate further, by assuming that the eigenvalue of a conserved operator $\hat{\mathcal{M}}$ (such as for example (\ref{eq_ConservedP})) is associated to the mass $M$ of the central black hole. 
Then, imposing that the inner product (\ref{eq_InnerQG}) of such an eigenstate with itself can be normalized to one
could imply a quantization of the eigenvalue $M$ into discrete mass levels $M_n$.
Just like the normalization condition implies discrete energy levels in the quantum description of a point particle.
However, at this stage, the possibility of a quantized mass spectrum is pure speculation.

For the purpose of this paper (uncertainty relations), we will restrict the discussion and assume 
the existence of a Hilbert space, an inner product, a reference state, an algebra of observables, vacuum expectation values $\langle {\mathcal{O}}\rangle$, and that operators can be written in the Heisenberg picture.
Still, it is also clear that  follow-up questions concerning the Schr\"odinger picture, the wave function $\Psi$, 
are certainly justified, interesting and deserve separate studies. 

%%%%%%%%%%%%%
\subsection{Gauge Fixing and Causal Structure}

When implementing the SER quantization, we brutally chose explicit coordinates and fixed the gauge of the field theory. Such a choice 
admittedly implies loss of generality and elegance. However, this shall not bother us as long as the mathematics is correctly describing the physical system of interest. Here, we have to be careful when we choose our observables. As discussed in subsection \ref{subsec_Obs}, not every real valued VEV of an operator is necessarily an observable. However, quantities which are actually directly linked to real observations, such as the ones entering geodesic motion, certainly fulfill this requirement.
Further, in the usual geometric dynamics for black holes~\cite{Kuchar:1994zk}, the author takes care of respecting the causal structure of the Schwarzschild spacetime represented in the Kruskal 
diagram and by introducing the Killing time.

In our case,
we choose a symmetry and fix the gauge. Thus, we start by the assumption that we have a static spacetime. The corresponding Killing symmetry exists and is built-in in the framework from the beginning by assuming that we have no time dependency in the metric-(operators). Since we integrate out time in our action, the corresponding integration constant can always be absorbed in the other integration constants $C_{ij}$.
However, it is interesting to note that the Killing symmetries provide a valuable tool when evaluating the motion of test particles in such QG-backgrounds in terms of q-desics~\cite{Koch:2025qzv}.

%https://www.lancaster.ac.uk/staff/schomeru/lecturenotes/Quantum%20Mechanics/S19.html

%%%%%%%%%%%%%
\subsection{The Cosmological Constant Problem}

When we started this project, we had \emph{no} intention to even mention this problem, but now, since the uncertainty relation for the metric operators $\hat g$ and $(\hat g \hat p^2)_W$ provided us with two bounds on this constant, we are obliged to address this topic.

The cosmological constant problem arises from the enormous discrepancy between the vacuum energy density predicted by quantum field theory and the tiny value required by cosmic acceleration measurements. 
As the most straightforward estimate, it
can be written as a dimensionless product of
the observed value of the cosmological constant $(\Lambda)_{obs}$ and the gravitational coupling constant:
\begin{equation}\label{eq_CCPobs}
\frac{\hbar }{c^3} \bigl(G\cdot  (\Lambda)_{obs}\bigr)=2\cdot 10^{-121}.
\end{equation}
The result is such a small dimensionless number that it marks one of the most severe fine-tuning puzzles in physics. While the cosmological constant dominates the current energy budget of the Universe, driving its accelerated expansion, its extreme smallness remains unexplained. Proposed resolutions span a wide range of ideas, yet no universally accepted solution has emerged
\cite{Sahni:2002kh,Prokopec:2006yh,Brodsky:2009zd,Kaloper:2013zca,Stojkovic:2013xcj,Padilla:2015aaa,Bass:2015yaa,Novikov:2016hrc,Nojiri:2016mlb,Wetterich:2017ixo,Hossenfelder:2018ikr,Padmanabhan:2006cj,Canales:2018tbn,Kanatchikov:2023gku}.

Since the uncertainty relation (\ref{eq_uncertfg0}), when evaluated at the Schwarzschild radius, provided us with a lower bound on the absolute value of $\Lambda$ for a given mass (or equivalently a lower bound on the allowed central mass for a given value of $\Lambda)$), let us evaluate the numerical value of this bound. For this, we use (\ref{eq_CCPobs}).
We find
\begin{equation}\label{eq_M0lower}
M_0 \;\ge\; \left( \frac{\tilde C_{0,1}}{\gamma^2}\right)^{1/3}\; 3\cdot 10^{32} \;\text{kg}.
\end{equation}
This is reminiscent of theoretical upper limit of compact objects known for neutron stars. In particular, if we consider the Tolman--Oppenheimer--Volkoff (TOV) limit, which is about $10^{31}~$kg, the numerical coincidence with (\ref{eq_M0lower}) might not be accidental:
The TOV limit arises from balancing the Pauli exclusion principle of particles against the gravitational pull of an object with mass $M_0$. 
Thus, the upper limit on such stars translates to a lower limit on the mass of black holes, as in (\ref{eq_M0lower}).
Furthermore, both the TOV limit and (\ref{eq_M0lower}) are essentially consequences of the uncertainty principle—one pertaining to particles in a spacetime background, and the other pertaining to spacetime itself. Thus, both limits rest on the same fundamental concept, though realized in very different ways.

Next, we apply the same reasoning to the upper limit (\ref{eq_LambdaUpper}), which was obtained from the large-radius expansion of (\ref{eq_uncertfg0}). 
The dimensionless inequality for the maximally allowed mass reads
\begin{equation}\label{eq_CCPCH}
M_0 \;\le\; c^2 \sqrt{\frac{3}{2\,G^2\,|\Lambda|}} \;=\; 1.4\cdot 10^{53}\;\text{kg}.
\end{equation}
This mass is remarkably close to the total mass of ordinary matter in the visible Universe, $M_0=1.5 \cdot 10^{53}~$kg. 
We find this relation between total mass and the value of $(\Lambda)$ rather surprising, since the measurement of $(\Lambda)_{obs}$ is based on observing the dynamical evolution of the Universe, while
our model arose from the quantization of GR imposing static 
and spherical symmetry with a single mass at the center. Note further that the inclusion of dark matter (if it contributes) would shift the value by a factor of five. Given the many orders of magnitude in (\ref{eq_CCPCH}) and the uncertainties involved, we should not be overly particular about this. 

Note that we can also invert equation (\ref{eq_CCPCH}) and state that the observed value of the cosmological constant (\ref{eq_CCPobs}) takes the maximal value allowed by the known observed mass in the Universe.
Before getting overly enthusiastic about this surprising result, we have to mention two factors which limit the universality and applicability of these interesting results:
\begin{itemize}
    \item First, we have to remember that this result is subject to the prior choice of a highly symmetric spacetime state, for which many integration constants vanish. There are three ways to gain better understanding of the physical role of the integration constants:
    \begin{enumerate}
        \item Study physical observables and thus identify the role of each constant $C_{i,j}$ in these measurable quantities.
        \item 
        Use the analogy in the non-relativistic quantum mechanics established in the appendix, and identify the meaning of a constant $C_{i,j}$ accordingly.
        \item 
        Develop a full description in terms of wave-functions in the Schr\"odinger picture, which then will allow to identify the role and skope of each constant $C_{i,j}$ in terms of the shape of the wave functions $\Psi(g)$.
    \end{enumerate}
    We are working all these approaches, but it will still take some time to make progress.
    \item 
    Our framework is time-invariant by construction, while the visible Universe with its $10^{53}$~kg is not. Which means that the mass-coincidence could also be just this, a coincidence.
\end{itemize}

%%%%%%%%%%%%%%%%%%%%
\section{Conclusion}
\label{sec_Concl}

In this work, we have developed a canonical quantization scheme for static, spherically symmetric spacetimes described by the Einstein--Hilbert action with a cosmological constant. Employing a reduced phase space approach, we quantized the gravitational degrees of freedom and demonstrated that the classical Schwarzschild--(Anti-)de Sitter solutions emerge in the semiclassical limit via the Ehrenfest theorem. All further information on the spacetime wavefunction $|\Psi\rangle$ is obtained in terms of integration constants of  Heisenberg's equations of motion for expectation values.

Beyond the recovery of classical dynamics, the quantized framework enabled us to explore nontrivial quantum features of the spacetime geometry. In particular, we derived uncertainty relations between geometric operators that arise naturally in the quantum theory. These relations suggest an intrinsic quantum indeterminacy of geometric quantities and reveal conceptual links to both generalized uncertainty principles and black hole thermodynamics. 

A notable outcome of our analysis is the existence of both lower and upper bounds on the mass parameter, depending on the value of the cosmological constant. 
Even though, we assumed a simple wave function with several vanishing integration constants for obtaining the upper mass bound, it is remarkable that when the cosmological constant is set to its observed value, the minimal mass predicted by our model is below the TOV limit, while the maximal mass closely approximates the total baryonic mass of the observable universe. This striking coincidence hints at a deep connection between quantum gravitational effects and large-scale cosmological structure.

Our findings provide a novel link to the observable effects of quantum gravity, at distance scales far larger than the Planck scale.
This was achieved by constraining the 
physically admissible spherically symmetric spacetimes. The presence of mass bounds also opens the door to intriguing phenomenological implications in astrophysics and cosmology.

Future directions include extending the quantization procedure to more general settings, such as charged or slowly rotating black holes, and exploring dynamical or cosmological scenarios. Our method, will further allow to test many more observables such as the parameters of geodesic motion. 

\textbf{Note added:}
After the article's first appearance on the arXiv, we became aware of closely related approaches:
\begin{itemize}
    \item One is formulated by Davidson and Yellin~\cite{Davidson:2012dt,Davidson:2014tda}. 
A difference to our results is that the authors find e.g. a constant expectation value for $\langle \hat g \rangle$, which is then adjusted by modifying the Hamiltonian to get a result which is in agreement with~(\ref{eq_gvev}).
\item The other is developed by Kanatchikov and collaborators, which does not require spherical symmetry. This formalism is called ``precanonical QG''~\cite{Kanatchikov:2013xmu,Kanatchikov:2014nua,Kanatchikov:2016neu} and it was further explored in the context of spin foams~\cite{Kanatchikov:2015kno}, quantum Yang-Mills theory~\cite{Kanatchikov:2018uoy,Kanatchikov:2017zfe}, teleparallel gravity~\cite{Kanatchikov:2023rqs}, and long-range observables~\cite{Kanatchikov:2025tlm}. 
\end{itemize}

%%%%%%%%%%
%\subsection{Before concluding}

%All shown steps are straight forward and all calculations were basically trivial. 
%Thus, the most likely scenarios are that all this has either been done a long time ago, or it is completely wrong, or a bit of both. Before submitting something like this to the arXiv, I prefer to ask people how certainly know this type of stuff much better than I do. Thus, don't be surprised when you find this in your inbox. Thanks for reading.

%%%%%%%%%%%%%%%%%
%\section{Acknowledgments}
\begin{acknowledgments}
%\section*{Acknowledgements}
We are  grateful to Maximo Ba\~nados for his class  on quantization of constrained systems, and to Daniel Grumiller for his valuable insights into the quantization of lower-dimensional gravitational models. We also acknowledge ChatGPT for its helpful discussions on Weyl ordering. Further appreciation goes to Thomas Thiemann for highlighting the relation to Kantowski--Sachs spacetimes, and to Robert Oeckl for offering comments on compositional quantum field theory and coherent states. Finally, We thank Renata Ferrero, Enrique Mu\~noz, Ren\'e Sedmik, Mario Pitschmann, Chad Briddon, Philipp Neckam, and Harald Skarke for their remarks and questions.\\

{\bf{Author Contributions: }}Conceptualization, B.K. and A.R.; methodology, B.K. and A.R.; validation, B.K. and A.R.; formal analysis, B.K. and A.R.; investigation, B.K. and A.R.; writing—original draft preparation, B.K.; writing—review and editing, B.K. and A.R.;
supervision, B.K; project administration, B.K; All authors have read and
agreed to the published version of the manuscript.\\

{\bf{Conflicts of Interest: }}The authors declare no conflict of interest.\\

{\bf{Data Availability Statement: }}There is no data available for this research\\

{\bf{Funding: }}This work was done without funding support

\end{acknowledgments}

%%%%%%%%%%%%%%%%%%%%%%%%%%%
\appendix

\section{The Free Particle}
\label{subsec_free_appendix}

In this subsection, we introduce some standard notation used in the SER approach. We revisit the non-relativistic point particle in quantum mechanics. All readers familiar with this textbook material may jump directly to the following section.
A free particle in one dimension is described by the action
\be
S= \int dt\, L = \int dt\, \frac{\dot{x}^2}{2m}.
\ee
From this formulation in terms of an action, we can switch to a Hamiltonian formulation by defining the conjugate momentum
\be
p = \frac{\partial L}{\partial \dot{x}},
\ee
and defining the Hamiltonian as
\be
H = \dot{x} p - L = \frac{p^2}{2m}. 
\ee
The conjugate variables $x$ and $p$ allow to define Poisson brackets
\be
\{A,B\} \equiv \frac{\partial A}{\partial x}\frac{\partial B}{\partial p} -
\frac{\partial B}{\partial x}\frac{\partial A}{\partial p}.
\ee
The Hamilton equations for a phase-space function $f$ are
\bea\label{eq_eomx_appendix}
\dot{f} &=& \{f, H\} + \frac{\partial f}{\partial t}.
\eea
Applying the standard rules of canonical quantization, we replace functions by operators 
$(H \rightarrow \hat{H}, \, p \rightarrow \hat{p} = i \hbar \frac{\partial}{\partial x}, \, \dots)$ and Poisson brackets by commutators, $\{A ,B\} \rightarrow \frac{1}{i\hbar}[\hat{A}, \hat{B}]$.\\
The Heisenberg equations of motion for an operator $\hat{A}$ then become
\be\label{eq_eomHeisi_appendix}
i \hbar \frac{d}{dt} \hat{A} = [\hat{A}, \hat{H}] + \frac{\partial \hat{A}}{\partial t}.
\ee
It is straightforward to write down and solve this equation for the expectation values of the operators $\hat{x},\; \hat{p},\; \hat{p} \hat{x} \hat{p},\; \hat{p}^2,\; \hat{x}^2,\; (\hat{x} \hat{p} + \hat{p} \hat{x})/2,\; \dots$.
The solutions read:
\bea\label{eq_xvevs1}
\langle \hat{1} \rangle &=& 1,\\
\langle \hat{p} \rangle &=& p_0, \\
\langle \hat{p}^2 \rangle &=& p_{2,0}, \\
\langle \hat{x} \rangle &=& \frac{p_0}{m} t + x_0, \\
\langle \hat{p} \hat{x} \hat{p} \rangle &=& \hbar^2 C_{1,2} + \frac{p_{2,0}}{m} t, \\
\left\langle \frac{\hat{x} \hat{p} + \hat{p} \hat{x}}{2} \right\rangle &=& \frac{p_{2,0}}{m} t + C_{1,1}, \\
\langle \hat{x}^2 \rangle &=& \frac{2}{m^2} \left( \frac{p_{2,0} t^2}{2} + m t C_{1,1} \right) + x_{2,0}.\label{eq_xvevsn}
\eea
Here, the integration constants $x_0,\, p_0,\, p_{2,0},\, C_{1,2},\, x_{2,0},\, C_{1,1}$ are real numbers, since the corresponding operators are Hermitian. Their values depend on the initial conditions of a given wave function.
For example, consider a wave function of some system in the Schr\"odinger picture which is in an energy eigenstate and parity symmetric
\be
\psi(x,t)=e^{i E t/\hbar} \psi(x),
\ee
with $\psi(x)=\psi(-x)$.
Then it is straight forward to show that the integration constant $p_0$ vanishes
\be\label{eq_symmetryExample}
\langle \hat p \rangle=i \hbar 
\int_{-\infty}^{+\infty} dx \; \psi(x) \psi'(x)=0,
\ee
due to symmetry reasons. Thus, the values of the integration constants in (\ref{eq_xvevs1}-\ref{eq_xvevsn}), depend i.a. on the symmetry properties of the wave function $\psi(x)$.
It is instructive to use the solutions (\ref{eq_xvevs1}-\ref{eq_xvevsn}) to verify the uncertainty principle:
\be\label{eq_UncertQM0_appendix}
\sigma_A^2 \sigma_B^2 =
\left(
\langle \hat{A}^2 \rangle - \langle \hat{A} \rangle^2
\right)
\left(
\langle \hat{B}^2 \rangle - \langle \hat{B} \rangle^2
\right)
\ge 
\left| \frac{1}{2i} \langle [\hat{A}, \hat{B}] \rangle \right|^2.
\ee
Using the operators $\hat{A} = \hat{x}$ and $\hat{B} = \hat{p}$, we find:
\be\label{eq_UncertQM1_appendix}
\sigma_x^2 \sigma_p^2 =
(p_{2,0} - p_0^2)
\left[
(x_{2,0} - x_0^2) - t \frac{2}{m} (p_0 x_0 - C_{1,1} )+ \frac{t^2}{m^2} (p_{2,0} - p_0^2)
\right]
\ge
\frac{\hbar^2}{4}.
\ee
We see that for large $t$, the left-hand side grows, which is consistent with the inequality. The minimal value of the left-hand side of equation (\ref{eq_UncertQM1_appendix}) is reached at $t=0$, where the uncertainty relation becomes:
\be\label{eq_UncertQM2_appendix}
(p_{2,0} - p_0^2)(x_{2,0} - x_0^2)
\ge
\frac{\hbar^2}{4}.
\ee
Thus, a perfect knowledge of the
position implies total ignorance of the momentum or vice versa. Further, we see from (\ref{eq_UncertQM1_appendix}) that there is no coherent state of a free particle in quantum mechanics. Even if the uncertainty relation is saturated at a given time $\sigma_x \sigma_p|_{t=t_c}=\hbar/2$, it will evolve away from such a state for later times.\\
With this, we have introduced all the quantum machinery necessary for this paper and proceed to our objective, the quantization of spherically symmetric spacetimes.
%%%%%%%%%%%%%%%%%%%%%%%%%%%
\section{Useful Relations}
Weyl operator ordering is a prescription for assigning a unique quantum operator to a corresponding classical function of position and momentum. It's especially useful when the classical quantities do not commute. The key property of Weyl ordering is that it makes the resulting quantum operator Hermitian if the classical function is real. The general formula for the Weyl ordered product of operators is given as follows:
\begin{equation*}
    (\hat{q}^n \hat{p}^m)_W = 
\sum_{k=0}^{\min(n,m)} 
\left(\frac{-i\hbar}{2}\right)^k  k!
\binom{n}{k}
\binom{m}{k}
\hat{q}^{\,n-k} \hat{p}^{\,m-k}.
\end{equation*}
Commutators between Weyl ordered products
of $\hat g$ and $\hat p$ can be written as sum of Weyl ordered products of lower rank.
We could not come up with a general formula, 
but here are two useful relations
\bea
\left[(\hat g^a \hat p^b)_W,\hat g \right]
&=&-i\hbar b (\hat g^a \hat p^{b-1})_W,\\
\left[(\hat g^a \hat p^b)_W,\hat p \right]
&=&i\hbar a (\hat g^{a-1} \hat p^{b})_W.
\eea
More formally, these can also be written as
\bea
\left[(\hat g^a \hat p^b)_W,\hat g \right]
&=&-i\hbar  \left(\frac{\partial(\hat g^a \hat p^{b})}{\partial \hat p}\right)_W,\\
\left[(\hat g^a \hat p^b)_W,\hat p \right]
&=&i\hbar  \left(\frac{\partial(\hat g^a \hat p^{b})}{\partial \hat g}\right)_W.
\eea

%%%%%%%%%%%%%%%%%%%%%%%%
\section{Expectation Values of Inverse Operators}
\label{sec_AppendB}

The circumference observable~\eqref{eq_ginvvev} depends on the expectation
value of the inverse metric operator, $\bigl\langle 1/\hat g \bigr\rangle$,
which in turn follows from the line element~\eqref{eq_LineElement}.
Assuming that it actually exists, how can we calculate it?
A convenient first step is therefore to evaluate the commutator
$[\hat H,\,\hat g^{-n}]$.  Substituting the result into the Heisenberg
equation yields
\begin{equation}\label{eq_Heisiginv}
\frac{d}{dr}\Bigl\langle \hat g^{-n}\Bigr\rangle
   = -\frac{n}{r}\!\left(1 - r^{2}\Lambda c^{2}\right)
     \Bigl\langle \hat g^{-(n+1)}\Bigr\rangle
     + \frac{n}{r}\,\Bigl\langle \hat g^{-n}\Bigr\rangle .
\end{equation}
A glance at this equation for~$n=1$ already exposes a difficulty
that persists for all~$n\ge 1$: the expectation value of a
\emph{lower} inverse power, such as
$\bigl\langle \hat g^{-1}\bigr\rangle$, depends on the expectation
value of the \emph{next higher} inverse power,
$\bigl\langle \hat g^{-2}\bigr\rangle$.  Attempting to solve the
hierarchy iteratively therefore forces us to determine an
\emph{infinite} tower of expectation values.
Due to cascading behavior we can not
find expectation values of inverse operators by the direct use of the Heisenberg equation of motion (\ref{eq_Heisiginv}).

Below, we will use a different strategy and show its utility by calculating the expectation values of $\langle \hat g^{-1}\rangle$ and $\langle \dot {\hat g}/\hat g\rangle$.
For this purpose we define the auxiliary quantity 
\be\hat U =1-\hat g,
\ee 
which is reminiscent of a potential operator.
With this auxiliary operator, we further can define the logarithm of the operator $\hat g$ as an infinite power series of $\hat U$
    \be\label{eq_ln}
    ln(\hat g)= - \sum_{n=1}^\infty\frac{\hat U^n}{n}.
    \ee
    Powers of $\hat U$ can now be written as binomials
    \be\label{eq_Delta}
    \hat U^n=\sum_{l=0}^{n} \left( \begin{array}{c}
         n  \\
         l 
    \end{array}\right)
    \hat g^l (-1)^{l}.
    \ee
    We realize that inserting (\ref{eq_Delta}) into (\ref{eq_ln}) we are able to express the logarithm of $\hat g$ as infinite sum of operators, whose expectation we already know.
Now we can act on (\ref{eq_ln}) with two different derivatives, to obtain the desired results $\langle \hat g^{-1}\rangle$ and $\langle \dot {\hat g}/\hat g\rangle$
\begin{itemize}
    \item $\langle \hat g^{-1}\rangle$\\
    If we apply the functional derivative $\tilde \delta$ with respect to $\hat g$ to (\ref{eq_ln}) and then evaluate the vacuum expectation value, we find
    \bea
    \left \langle\frac{1}{\hat g} \right \rangle = \left \langle\frac{\tilde \delta \ln (\hat g)}{\tilde \delta \hat g} \right \rangle  &=&
    \sum_{n=1}^\infty
     \left \langle
     \hat U^{n-1}
     \right \rangle\\ &=& 3-3\langle \hat g\rangle+\langle \hat g^2\rangle+ {\mathcal{O}}\left(\left \langle
     \hat U^3
     \right \rangle\right),\label{eq_ginv1PN}
    \eea   

where in the last line we truncated the sum and only retained the leading and subleading order in $\hat U$. In classical general relativity, this would correspond to a first Post-Newtonian expansion.
\item 
$\langle \dot {\hat g}/\hat g\rangle$\\
If we act with a radial derivative 
on (\ref{eq_ln}) and subsequently use the Heisenberg equation (\ref{eq_Ehrenfest}) for the expectation value of the operators 
we find
    \bea
    \label{eq_gprime/g_vev}
    \langle \hat g'/\hat g \rangle =
    \sum_{n=1}^\infty 
    \langle \hat U^{n-1} \hat g' \rangle
    = \sum_{n=1}^\infty 
    \langle(1 - \hat g)^{n-1} \frac{(1- r^2 \Lambda - \hat g )}{r} \rangle =  \langle \frac{1- r^2 \Lambda - \hat g }{r} \rangle + \mathcal{O}(\langle \hat U^1 \rangle) \approx \frac{1- r^2 \Lambda }{r} - \frac{\langle \hat g \rangle}{r}. 
    \eea
    This expression exclusively contains  positive powers of $\hat g$, thus we can insert for each term the corresponding vacuum expectation value (\ref{eq_VevGeneral}).
    The resulting expression contains an infinite sum of integration constants $C_{n,0}$. However, we always can calculate a Newtonian expansion, just like we did in (\ref{eq_ginv1PN}). To leading order this gives
    \bea
   \langle \hat g'/\hat g \rangle = \frac{1- r^2 \Lambda }{r} - \frac{1}{r} \left( (1+\epsilon_{0,0})+\frac{-\frac{2GM_{0}}{c^{2}}\bigl(1+\epsilon_{1,0}\bigr)}{r}-\frac{r^2 \Lambda(1+\epsilon_{0,0})}{3} \right)  + \mathcal{O}(\langle \hat U^1 \rangle)
    \eea

    Another interesting scenario is to assume that  the spacetime state is an eigenfunction of the operator $\hat g$ , such that $\hat g| \Psi_g\rangle=\langle \hat g \rangle \,| \Psi_g\rangle$. In this case, the integration constants are not independent and we can use
    $\langle \Psi_g|\hat g^l|\Psi_g\rangle= \langle \hat g\rangle^l$ or equivalently
    $C_{l,0}=C_{1,0}^l$.
    Therefore, if we assume such a state $| \Psi_g\rangle$, we can perform   the entire sum
    over $n$ and write
    \be\label{eq_sumsumsum2}
     \langle \Psi_g| \hat g'/\hat g |\Psi_g\rangle=\frac{
     1-\langle \hat g\rangle+r^2 \Lambda
     }{\langle \hat g\rangle^2 r}.
    \ee
%    \textcolor{red}{??? We readily confirm that in the weak curvature limit and for negligible $\Lambda$
%    this result is consistent with (\ref{eq_PNApprox}). Missing equation ??? }

\end{itemize}
\section{Possible interpretation of the integration constants in  $[\hat g, \hat p]$}

Let's now discuss several aspects of the uncertainty relation (\ref{eq_uncertGR0}) by comparing to different results in the literature. For the sake of this comparison let's use units with $c=1$
\begin{itemize}
    \item First, we compare it to the uncertainty relation of spacetime which was previously suggested in~\cite{Sasakura:1999xp}
\be\label{eq_uncertSas}
\frac{(\delta V)^2}{(\delta Z)^2 G^2}\ge \frac{\hbar^2}{4}.
\ee
Here, $\delta V\sim (\delta l)^3$ stands for a characteristic volume and $Z$ stands for a characteristic time.
In our case~(\ref{eq_uncertGR0}), the first parenthesis stands for the square of a characteristic length  and the second parenthesis stands for the square of some energy whose units are given from the definition (\ref{eq_GammaDef}). Inserting this definition
we can write (\ref{eq_uncertGR0}) in the same units and notation as~(\ref{eq_uncertSas})
\be\label{eq_uncerComp}
\frac{\delta Z^2 (\delta l)^2}{G^2}\ge \frac{\hbar^2}{4}.
\ee
We realize that  (\ref{eq_uncertSas}) and (\ref{eq_uncerComp}) agree if we identify the characteristic length scale with the characteristic time scale $\delta Z \sim \delta t$. Doing this identification, the statement basically reduces to $(\delta l)^2/G\ge \hbar$, which also could be guessed from a pure dimensional analysis. It is also worth noting that we can reformulate the inequality and write it as $(\delta l)^2 \geq \hbar G/c^3 = (\delta l_{Pl})^2$.
\item
If one variance in (\ref{eq_uncertGR0}) vanishes, the other has to diverge,
to keep the uncertainty relation intact.
Let's see what happens if the uncertainty of the momentum operator vanishes  $\sigma_p^2\rightarrow 0$.
This happens when the two energy scales coincide $\sqrt{C_{0,2}}=\Gamma$ and $ C_{0,1}=M_X c^2$. 
Now we use this energy in the definition (\ref{eq_GammaDef})
and invoke the typical relation 
between time scales ($Z$) and inverse temperature 
\be\label{eq_BringTemp}
Z=\frac{\hbar}{4 \pi k_B} \frac{1}{T}.
\ee
With this, we find an upper limit on the temperature associated to a given central mass scale 
\be\label{eq_THawk}
T\le \frac{\hbar}{8 \pi  G k_B M_X}.
\ee
% omitted c^3  in numerator
Thus, the saturation of the $\hat g$ $\hat p$ uncertainty
yields the Hawking temperature of the mass $M_X$ as upper bound. In Ref.~\cite{Pinochet:2018ysz}, the uncertainty principle was investigated to obtain the Hawking temperature. Our SER quantization approach also allows us to obtain the Hawking temperature from the uncertainty relations between the metric operators and the calculation of the expectation values. It is worth to keep in mind that, as in quantum/statistical physics, using a wick rotation $t \rightarrow -i t $, one can switch between Minkowski and Euclidean signature. This allows us to make sense of such interpretation since we work outside of the event horizon of the black hole and the element does not have terms such as $dtdr$.
%Therefore, our approach allows a deeper understanding of why it is possible to obtain the Hawking temperature from heuristic arguments based on the uncertainty principle~\cite{Pinochet:2018ysz}.
\end{itemize}

\section{Complementary material on the operators}
\label{sec_AppendD}
This work introduced a quantization of the minisuperspace of spherically symmetric static spacetime where the metric operators, considered as the canonical variables, are promoted to operators. Even though a pure operator formalism is not necessary for the scope of this work, briefly provide some complementary material for the reader on 
the domain of the operators, their self adjointness, and a discussion on the measure.\\
We first look at the symmetry properties of the operators $\hat g$ and $\hat p$. We set the measure $d\mu (g) = d g$ which is the typical Lebesgue measure.
\begin{equation}
    \langle \psi ( \hat g ) | \phi ( \hat g) \rangle = \int d \mu ( \hat g) \psi ( \hat g )^* \phi ( \hat g) = \int d g  \psi ( \hat g )^* \phi ( \hat g) .
\end{equation}
We want to show that
\begin{equation}
    \label{eq_symH_v1}
    \langle \hat g \psi ( \hat g ) | \psi ( \hat g) \rangle - \langle \psi ( \hat g ) | \hat g \psi ( \hat g) \rangle = 0
\end{equation}
and
\begin{equation}
    \label{eq_symP_v1}
    \langle \hat p \psi ( \hat g ) | \psi ( \hat g) \rangle - \langle \psi ( \hat g ) | \hat p \psi ( \hat g) \rangle = 0.
\end{equation}
The metric operator $\hat g$ is  trivially symmetric, confirming (\ref{eq_symH_v1}). For the momentum operator $\hat p $ one has to integrate by parts
\begin{equation}
   \langle \hat p \psi ( \hat g ) | \psi ( \hat g) \rangle - \langle \psi ( \hat g ) | \hat p \psi ( \hat g) \rangle  =  i\hbar [ \psi^* \psi ]_{Boundary} .
\end{equation}
Further assuming a compact support for the wave functions at the boundaries one gets $ i\hbar [ \psi^* \psi ]_{Boundary}= 0$, confirming (\ref{eq_symP_v1}). \\
Next we discuss the domain of the operators and their self adjointness.\\
\newline
We define the metric operator as $\hat g: D(g) \rightarrow \mathcal{H}$ and $\hat g \psi(g) = g \psi(g) \, \forall \psi \in \mathcal{H}$ with the domain $D(g)=\{ \psi, g\psi \in \mathcal{H} | \int dg\, g^2 \psi^2 < \infty \}$.\\
%Its adjoint extension is defined as $\hat g^{\dagger}:\; D(g^{\dagger}) \rightarrow \mathcal{H}$ with the expression $\hat g^{\dagger} \psi(g) = g \psi(g) \quad \forall \psi \in \mathcal{H}$.
Since $\hat g$ is symmetric, we have $\hat g \subset \hat g^{\dagger}$. In order to show the operator is self adjoint, we have to show $D(g) = D( g^{\dagger})$. \\
Definition: $\forall \psi \in D(g)$ and $\forall \phi \in D(g^{\dagger})$, $\exists \eta \in \mathcal{H}$ such that $\langle \phi ,  \hat  g \psi \rangle = \langle \eta , \psi \rangle $ is satisfied. We compute the integral and then define the adjoint of $\hat g$ by setting $ g\phi=\eta= \hat g^{\dagger} \phi$.
Doing so, we find the domain of $\hat g^{\dagger}$ to be $D(g^{\dagger})= \{ \phi, g\phi \in \mathcal{H} | \int dg g^2 \phi^2 < \infty \}$ which is equal to the domain of $\hat g$. Thus, we have $D(g)= D(g^{\dagger}) \rightarrow g= g^{\dagger}$.\\
\newline
The situation for the momentum operator is not trivial and the operator is not self adjoint in general. However, defining the momentum operator in Sobolev space (1,2) of $ W^{1,2}=H^1=\{\psi \in \mathcal{C}^{1}_c (\mathbb{R} )\;\; |\;\; \psi \;\; \text{being weakly derivable} \;\; \} $ one can show that the momentum operator $\hat p$ has a unique self adjoint extension. One can generalize the Sobolev space to $H^{\infty}$ where $\psi \in \mathcal{C}^{\infty}_c$.\\
The momentum operator is symmetric and we have $\hat p \subset \hat p^{\dagger}$. We need to show $D(p)= D(p^{\dagger})$.\\
%The adjoint extension of $\hat p$ is defined as $\hat p^{\dagger}:D(p^{\dagger}) \rightarrow \mathcal{H}$ with the expression $\hat p^{\dagger} \psi(g) = - i \hbar \partial_g \psi(g) \;\; \forall \psi \in D(p^{\dagger})$.\\
Definition: $\forall \psi \in D(p)$ and $\forall \phi \in D(p^{\dagger})$, $\exists \eta \in \mathcal{H}$ such that $\langle \phi , \hat p \psi \rangle = \langle \eta , \psi \rangle $ is satisfied. 
\begin{equation}
    \langle \phi , \hat p \psi \rangle = \int \phi^* p \psi = -i \hbar \int dg \phi^* \partial_g \psi = - i\hbar \{ \phi^* \psi 
    \}^{\infty}_{- \infty}+ i \hbar \int dg \partial_g \phi^* \psi =  \int dg \left(- i \hbar \partial_g \phi \right)^* \psi.
\end{equation}
The boundary term vanishes due to compact support. We identify $\eta = -i \hbar \partial_g \phi = \hat p^{\dagger} \phi $ since $\phi \in D(p^{\dagger})$. We find the following domain for the adjoint extension of the momentum operator, $D(p^{\dagger})= H^1=\{\phi \in \mathcal{C}^{1}_c (\mathbb{R} )\;\; |\;\; \phi \; \text{being weakly derivable} \; \} $ and thus $D(p) = D(p^{\dagger}) \rightarrow p = p^{\dagger}$ on the Sobolev space ($1\geq$,2).

\begin{center}
{\bf{References}}
\end{center}
\bibliography{ref.bib}

\end{document}